\documentclass[10pt,journal]{IEEEtran}
\usepackage{color}
\usepackage{epsfig}
\usepackage{amsmath, amsthm}
\usepackage{mathtools}  
\mathtoolsset{showonlyrefs}
\usepackage{amssymb}
\usepackage{amscd}
\usepackage{threeparttable}
\usepackage{graphicx}
\usepackage{subfigure}

\begin{document}
\title{Controlador LQR y SMC Aplicado a Plataformas Pendulares}
\author{Miguel F. Arevalo-Castiblanco, C. H. Rodriguez-Garavito, Álvaro A.Patiño-Forero, José F. Salazar-Cáceres}

\maketitle

\begin{abstract}
Una plataforma pendular es una estructura  robótica comúnmente empleada en el diseño de controladores dada su dinámica no lineal; este trabajo presenta el modelamiento, diseño e implementación de un controlador óptimo LQR y un controlador en modo deslizante SMC aplicado a dos plataformas comerciales, el péndulo rotatorio invertido de Quanser (RotPen) y el péndulo móvil de Lego (NxtWay). El aporte de este trabajo es presentar una metodología de implementación de controladores sobre plataformas pendulares, atendiendo las respectivas restricciones de hardware y software en prototipos comerciales. El artículo presenta el comportamiento de los controladores diseñados sobre el modelo analítico comparado con su implementación real.
\end{abstract}

\section{Introduction} \label{S1}
Es común encontrar aplicaciones en donde se emplean  estructuras pendulares con aplicación en robots humanoides, el auge de estas aplicaciones ha incentivado a la comunidad científica a plantearse nuevas investigaciones en el campo del modelamiento y el control, con el fin de optimizar los procesos en donde estas estructuras se encuentran presentes \cite{Li}.

Una de las estructuras robóticas móviles que ha tenido mayor acogida en los últimos años es el péndulo invertido de dos ruedas, con plataformas comerciales como el \textit{Segway} o el \textit{Hoverboard}; la investigación en el área del control ha abordado este tipo de plataformas con mucho interés, dada su fácil construcción y su comportamiento dinámico no lineal, junto con su conocida inestabilidad. Recientemente, varios son los fabricantes que desarrollan esta clase de estructuras, con fines académicos o industriales, el fabricante Quanser con sus plataformas de péndulo rotatorio invertido y doble péndulo rotatorio invertido, o el propio fabricante Hoverboard con las plataformas del mismo nombre, son algunos de los ejemplos conocidos que se encuentran comercialmente \cite{Hoverboard, Quanser}.

El objetivo de funcionamiento de estas estructuras pendulares, es lograr la estabilización del péndulo alrededor del punto de equilibrio con la menor energía invertida en su actuación \cite{Canale}; es común encontrar diferentes autores que desarrollan controladores para estructuras pendulares bajo diferentes metodologías, algunos de los casos documentados serán descritos a continuación.

Los primeros controles que se implementaron en péndulos invertidos, fueron lazos de control clásico, tipo PID o realimentación de estados, los cuales, al ser diseñados sin criterios óptimos, generan ineficiencia y restringen su aplicación a posiciones o condiciones específicas \cite{Prakash}. Con la ampliación de las teorías de control modernas, el uso de controladores óptimos aumentaron la eficiencia de diferentes sistemas, incluyendo las estructuras pendulares; el controlador óptimo más utilizado en la literatura es el controlador LQR (Linear Quadratic Regulator), este pondera heurísticamente las variables de estado de un sistema \cite{Murcia}, y usa constantes de realimentación que minimizan al máximo el error presente en las señales controladas \cite{Tijani}; es común que las variables a minimizar cuenten con un componente de ruido coloreado, que afecta el rendimiento del controlador; por esta razón, de manera alternativa se desarrolla el controlador LQG (Linear Quadratic Gaussian), en donde a partir del diseño de un Filtro de Kalman, se minimiza este ruido hasta el limite del ruido blanco, y se mejora el rendimiento del sistema \cite{Prakash}. Estas características han hecho que esta técnica de control sea una de las más utilizadas y documentadas para estructuras pendulares, sin embargo, no es la única metodología eficiente para ser implementada en péndulos; diferentes autores a lo largo de los últimos años han desarrollado nuevas técnicas de control que permiten mejorar las aplicaciones en donde se involucren esta clase de estructuras \cite{Mohan, Al-Jodah, Barya}. Entre las técnicas más utilizadas de manera alternativa a los controladores LQR se encuentran técnicas como el control en modo deslizante (Sliding Mode Control-SMC), en el cual se escoge una superficie de deslizamiento según el comportamiento de las variables de estado, y se seleccionan las constantes de realimentación que permitirán que las variables tomen la trayectoria definida de forma estable, a través de los criterios de estabilidad de Lyapunov \cite{Chakraborty} \cite{Portilla}.

Es común encontrar también controladores de carácter predictivo, en donde a partir de un modelo completo del sistema pendular, es posible modelar sus perturbaciones, para que el controlador sea capaz de rechazarlas al momento que sucedan; es una estrategia de control eficiente debido a que no solo tiene en cuenta el modelamiento de la planta, sino que además incluye una definición analítica de las perturbaciones que pueden intervenir en el modelo; esta estrategia requiere un pleno conocimiento del modelo dinámico no lineal y linealizado del sistema y sus perturbaciones \cite{Canale}\cite{Fontanet}; tarea que en ocasiones suele ser más complicada que el diseño de un controlador en estructuras pendulares \cite{Roman}.

Por último, la técnica que ha tenido más desarrollo a lo largo de los últimos años, ha sido la relacionada con el control robusto ($H_2/H_\infty$) \cite{Barya}; esta técnica maneja intervalos de tolerancia para las variables de estado, los cuales permiten que el sistema tenga un comportamiento más amplio y con inmunidad a cambios no drásticos, sin embargo la dificultad matemática que representa este tipo de controladores, hace que su implementación posea un grado de complejidad elevado, que dificulta su ejecución en sistemas de procesamiento de datos. \cite{Al-Jodah}.

Las diferentes técnicas desarrolladas en los últimos años para el control de estructuras pendulares, ha permitido una mayor eficiencia en sus realizaciones, trayendo consigo, un incremento en su uso comercial \cite{Canale}; es así como es posible observar en la industria diversas aplicaciones en sectores como: aeroespacial, biomecánica o transporte; con usos como el control activo para despegue de cohetes \cite{Paryathy}, la estabilidad de prototipos bípedos caminantes \cite{daSilva} o el mencionado medio de transporte Segway \cite{Cruz}.

El objetivo de la mayoría de los trabajos relacionados, se enfoca en el desarrollo de las metodologías de control en un único prototipo \cite{Al-Jodah, Barya, Chakraborty, Li, Mohan, Murcia, Prakash}; sin embargo, un análisis comparativo del comportamiento de las técnicas de control sobre diferentes plantas permite evaluar la factibilidad de la implementación  de controladores, teniendo en cuenta restricciones físicas como la facilidad de medida de las variables de estado o el costo computacional relacionado con la velocidad de procesamiento del hardware donde se embebe el controlador. El enfoque de este trabajo es llevar a cabo la comparación de las técnicas de control LQR y SMC en dos diferentes prototipos, un péndulo rotatorio invertido de Quanser (RotPen), y una péndulo invertido móvil de Lego (NxtWay), buscando que los resultados del control sean genéricos y sea posible hacer una comparación, validando su implementación en diferentes plantas ante una misma metodología de control.

Cada uno de los pasos de la metodología se presenta en las siguientes secciones así: la sección II hace énfasis en el modelo dinámico de ambas estructuras, posteriormente en la sección III, se tomará este modelo, y a partir de su linealización y su representación en espacio de estados, se realiza el cálculo de cada uno de los controladores; en la sección IV estos controladores serán implementados en los respectivos entornos, y por último en la sección V, se validará y comparará el rendimiento de los controladores bajo índices de calidad.

\section{Modelo Dinámico}
\begin{figure}[htbp]
	\centering
	\subfigure[Vista frontal de la plataforma]{
		\includegraphics[width=0.23\textwidth]{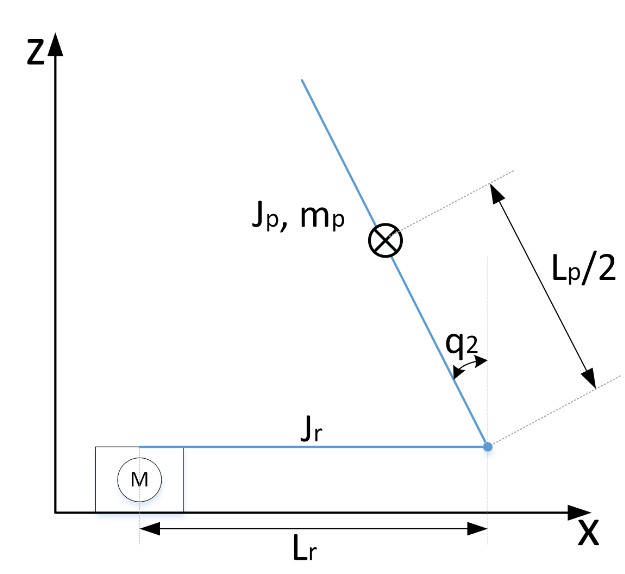}}
	\subfigure[Vista superior de la plataforma]{
		\includegraphics[width=0.23\textwidth]{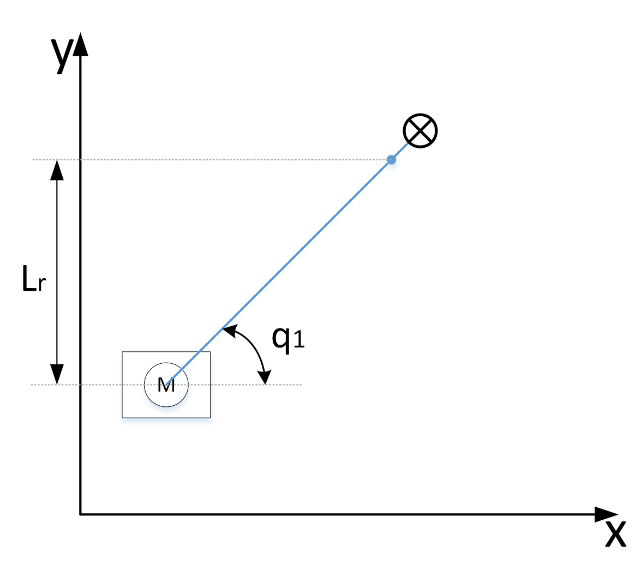}}
	\caption{Esquema del péndulo rotatorio invertido de Quanser, la M mostrada en ambas figuras corresponde al servomotor empleado.}
	\label{fig:fig1}
\end{figure}

\begin{figure}[htbp]
	\centering
	\subfigure[Vista frontal de la plataforma]{
		\includegraphics[width=0.23\textwidth]{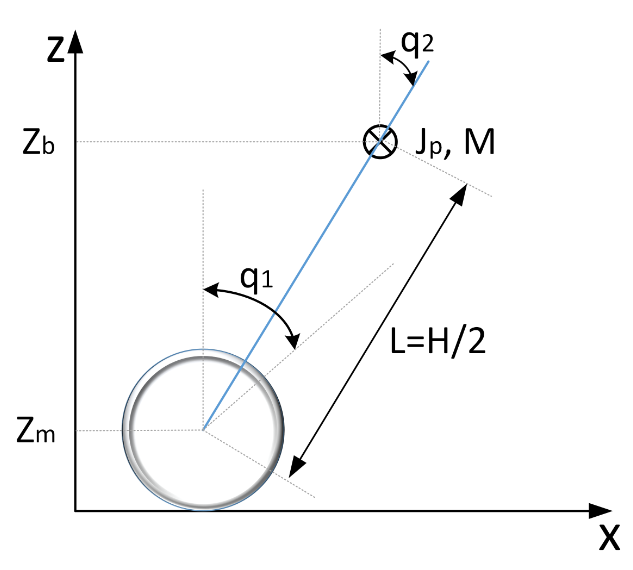}}
	\subfigure[Vista superior de la plataforma]{
		\includegraphics[width=0.23\textwidth]{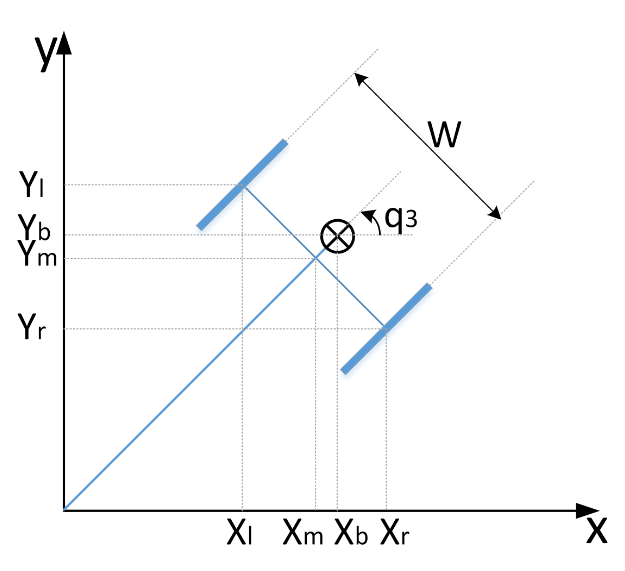}}
	\caption{Esquema del péndulo invertido móvil NxtWay}
	\label{fig:fig2}
\end{figure}

Para el control de un sistema, es necesario tener un conocimiento pleno de su modelo dinámico; como se menciona en la sección anterior, las plataformas a analizar son el péndulo rotatorio invertido de Quanser (RotPen) y el péndulo móvil de Lego (NxtWay), en las figuras \ref{fig:fig1} y \ref{fig:fig2} se observan las estructuras base de cada plataforma, las cuales definen las posiciones y velocidades que serán utilizadas como variables de estado en sus respectivos modelamientos.

El modelo dinámico en ambos casos se obtiene a partir de un método basado en energía, que parte de la definición del operador lagrangiano \cite{Kelly}; es importante aclarar que las energías surgen a partir del análisis de dos eslabones físicos para el modelo RotPen ([$q_1,q_2$]), como se observa en la figura \ref{fig:fig1} y para la plataforma NxtWay mostrada en la figura \ref{fig:fig2} se presenta un movimiento libre de dos grados de libertad definido por su posición $q_1$ y orientación respecto a z, $q_3$ que para propósitos de simplificación se hará igual a cero, y por la posición del polo $q_2$. El listado de los parámetros utilizados para el modelamiento de cada estructura se observa en las tablas \ref{table:tabla1} y \ref{table:tabla2}.

El objetivo de cada modelamiento es llegar a la definición de la ecuación de la robótica \cite{Barrientos}; en cada caso la entrada será el voltaje de los actuadores; su definición matricial se presenta en la ecuación \eqref{eq3}; en las ecuaciones  \eqref{eq:Gquanser} y \eqref{q:Gnxtway} se especifican cada uno de los términos involucrados en esta ecuación para el RotPen y el NxtWay respectivamente.

\begin{equation}
M_x\left(q\right) \ddot q + C_x\left(q, \dot q\right) \dot q + G_x\left(q\right) = V_x
\label{eq3}
\end{equation}

donde para el RotPen las matrices se representan como:

\begin{equation*}
M_Q = 
\left[\begin{array}{cc}
\scalebox{0.65}{$\gamma(m_r{\left(\frac{L_r}{2}\right)}^2+m_pL_r^2+J_r+m_p{\left(\frac{L_p}{2}\right)}^2s^2\left(q_2\right))$}& 
\scalebox{0.65}{$-\frac{1}{2}m_pL_pL_rc\left(q_2\right)\gamma$}\\
\scalebox{0.65}{$-\frac{1}{2}m_pL_pL_rc\left(q_2\right)$}&
\scalebox{0.65}{$m_p{\left(\frac{L_p}{2}\right)}^2+J_p$}
\end{array}\right]
\label{eq:Mquanser}
\end{equation*}
\begin{equation*}
C_Q = 
\left[\begin{array}{cc}
\scalebox{0.65}{$2s\left(q_2\right)c\left(q_2\right)m_p{\left(\frac{L_p}{2}\right)}^2\gamma\dot{q_2}+\gamma{f_r}+K_mK_g$}& 
\scalebox{0.65}{$\frac{1}{2}m_pL_pL_rs\left(q_2\right)\dot{q_2}\gamma$}\\
\scalebox{0.65}{$-\dot{q_1}m_p{\left(\frac{L_p}{2}\right)}^2s\left(q_2\right)c\left(q_2\right)$}&
\scalebox{0.65}{$f_p$}
\end{array}\right]
\label{eq:Cquanser}
\end{equation*}
\begin{equation}
G_Q = 
\left[\begin{array}{cc}
\scalebox{0.65}{$0$}&\\
\scalebox{0.65}{$-\frac{L_p}{2}m_pgsin\left(q_2\right)\frac{R_m}{K_mK_g}$}
\end{array}\right]\\
\label{eq:Gquanser}
\end{equation}

Y para el NxtWay como:

\begin{equation*}
M_N = \frac{1}{\alpha}
\left[\begin{array}{cc}
\scalebox{0.75}{$2mR^2+MR^2+2J_w+{2n}^2J_m$}& 
\scalebox{0.75}{$MLRc(q_2)-2n^2J_m$}\\
\scalebox{0.75}{$-(MLRc(q_2)-2n^2J_m)$}&
\scalebox{0.75}{$-(ML^2+J_{q_2}+2n^2J_m)$}
\end{array}\right]
\label{eq:Mnxtway}
\end{equation*}
\begin{equation*}
C_N = \frac{1}{\alpha}
\left[\begin{array}{cc}
\scalebox{0.75}{$2\left(\beta+f_w\right)$}& 
\scalebox{0.75}{$-2\beta-MLR\dot{q_2}s(q_2)$}\\
\scalebox{0.75}{$2\beta$}&
\scalebox{0.75}{$-2\beta$}
\end{array}\right]
\label{q:Cnxtway}
\end{equation*}
\begin{equation}
G_N = \frac{1}{\alpha}
\left[\begin{array}{c}
\scalebox{0.75}{$0$}\\
\scalebox{0.75}{$MgLs(q_2)$}
\end{array}\right]
\label{q:Gnxtway}
\end{equation}

Las matrices obtenidas definen el comportamiento no lineal de cada uno de los sistemas; para un correcto funcionamiento de los controladores implementados, es necesaria la linealización de los sistemas en un punto de operación estable, esta linealización es llevada a cabo mediante series de Taylor; y su resultado toma la forma de un modelo en variable de estado, como se muestra en la ecuación \eqref{eq4} y \eqref{eq4_1}.

\begin{equation}
\dot{x}(t)=A_xx(t)+B_xu(t)
\label{eq4}
\end{equation}
\begin{equation}
y(t)=C_xx(t)+D_xu(t)
\label{eq4_1}
\end{equation}

El vector de estados de cada plataforma esta compuesto a partir de las posiciones mostradas en las figuras \ref{fig:fig1} y \ref{fig:fig2} y sus respectivas velocidades, $x(t)=[q_1,q_2,\dot{q_1},\dot{q_2}]$; los valores de cada matriz para cada estructura se encuentran definidos en las tablas \ref{table:tabla5} y \ref{table:tabla6}; los valores de las matrices C y D, se encuentran definidas de manera estándar.

\begin{table}[htbp]
	\begin{center}
		\caption{Parámetros físicos RotPen. \cite{Quanser}}
		\label{table:tabla1}		
		\begin{tabular}{|c|c|p{4cm}|}
			\hline
			Parámetro & Und. & Descripción \\
			\hline \hline
			g=9.81 & $\frac{m}{s^2}$ & Gravedad \\ \hline
			$m_p$=0.127 & kg & Masa del péndulo \\ \hline
			$L_p$=0.337 & m & Longitud total del péndulo \\ \hline
			$J_p$=0.0012 & kg$m^2$ & Inercia del péndulo \\ \hline
			$m_{r}$=0.257 & kg & Masa del brazo rotatorio \\ \hline
			$L_r$=0.216 & m & Longitud del brazo rotatorio \\ \hline
			$J_{r}$=$9.98x10^{-3}$ & kg$m^2$ & Inercia del brazo\\ \hline
			$f_p$=0.0024 & $\frac{Nms}{rad}$ & fricción brazo - péndulo  \\ \hline			
			$f_r$=0.0024 & $\frac{Nms}{rad}$ & Fricción motor - brazo \\ \hline
			$R_{m}$=2.6 & $\Omega$ & Resistencia del motor CC \\ \hline
			$L_{m}$=0.18 & $mH$ & Inductancia del motor CC \\ \hline
			$K_m$=0.00767 &  $V\frac{s}{rad}$ & Constante Contra electromotriz motor CC \\ \hline
			$K_t$=0.00767 &  $\frac{Nm}{A}$ & Constante torque del motor \\ \hline
			$\eta_{g}$=0.9 & $ $ & Eficiencia caja reductora \\ \hline
			$\eta_{m}$=0.69 & $ $ & Eficiencia del motor \\ \hline
			$K_{enc}$=4096 & $\frac{conteos}{rev}$ & Resolución encoder \\ \hline
			$K_g$=70 & & Relación caja de engranajes\\ \hline
			$\gamma = \frac{R_m}{K_tK_g\eta_{g}\eta_{m}}$ & $ $ & Constante del motor \\ \hline
			$V_m$=6 & $V$ & Voltaje nominal del motor \\ \hline						
		\end{tabular}
	\end{center}
\end{table}

\begin{table}[htbp]
	\begin{center}
		\caption{Parámetros físicos NxtWay. \cite{Yamamoto}}
		\label{table:tabla2}		
		\begin{tabular}{|c|c|p{4cm}|}
			\hline
			Parámetro & Und. & Descripción \\
			\hline \hline
			g=9.81 & $\frac{m}{s^2}$ & Gravedad \\ \hline
			m=0.03 & kg & Peso de la rueda \\ \hline
			R=0.02 & m & Radio de la rueda \\ \hline
			$J_W=\frac{mR^2}{2}$ & kg$m^2$ & Inercia de la rueda \\ \hline
			M=0.6 & kg & Peso del cuerpo \\ \hline
			W=0.14 & m & Ancho del cuerpo \\ \hline
			D=0.04 & m & Profundidad del cuerpo \\ \hline
			H=0.27 & m & Altura del cuerpo \\ \hline
			L=0.12 & m & Distancia centro masa eje rueda \\ \hline
			J$q_2=\frac{ML^2}{3}$ & kg$m^2$ & Momento de inercia pitch \\ \hline
			J$q_3=\frac{M(W^2+D^2)}{12}$ & kg$m^2$ & Momento de inercia yaw\\ \hline
			$J_m=1x10^{-5}$ & kg$m^2$ & Momento de inercia motor \\ \hline
			$f_m$=0.0022 & $\frac{Nms}{rad}$ & Fricción dinámica cuerpo-motor \\ \hline
			$f_W$=0 &  $\frac{Nms}{rad}$  & Fricción dinámica rueda-suelo \\ \hline			
			$R_m$=6.69 & $\Omega$ & Resistencia del motor CC \\ \hline
			$K_b$=0.468 & $V\frac{s}{rad}$ & Constante Contra electromotriz motor CC \\ \hline
			$K_t$=0.317 & $\frac{Nm}{A}$ & Constante torque del motor \\ \hline
			$\eta=1$ &  & Eficiencia de transmisión \\ \hline
			$\alpha=\frac{nK_t}{R_m}$ & $ $  & Constante del motor \\ \hline
			$\beta=(\frac{nK_tK_b}{R_m})+f_m$ & $ $ & Constante del motor \\ \hline
			$V_m$=10 & $V$ & Voltaje nominal del motor \\ \hline				
		\end{tabular}
	\end{center}
\end{table}

\section{T\'ecnicas de Control}

En esta sección se presenta el diseño de dos técnicas de control moderno para validar la metodología propuesta: Control óptimo a partir de un regulador cuadrático lineal (LQR) y un Control en modo deslizante (SMC).

\subsection{Control LQR}

Con las matrices de estado, es posible realizar los cálculos para los controladores óptimos LQR; esta técnica de control parte del modelo dinámico de cada sistema, para obtener una matriz de realimentación que minimice un indice de calidad energético dado en \cite{Murcia}, como se muestra en la ecuación \eqref{eq5}.

En el caso del NxtWay, se emplean dos lazos de control, un lazo proporcional integrativo, PI, para la posición del móvil $q_1$ y un segundo lazo de control LQR para el equilibrio del péndulo, donde interviene parte del vector de estado $[q_2, \dot{q_1}, \dot{q_2}]$, la  separación del espacio de estados se puede efectuar gracias al desacople dinámico de $q_1$ en las ecuaciones de aceleración $\ddot{q_1}$ y $\ddot{q_2}$; el resultado de este análisis se evidencian en la tabla \ref{table:tabla6}.

\begin{equation}
J=\int_0^{\infty{}}\left(x^TQx+u^TRu\right)dt
\label{eq5}
\end{equation}

\begin{table}[htbp]
	\begin{center}
		\caption{Parámetros matrices de estado plataforma RotPen}
		\label{table:tabla5}		
		\begin{tabular}{|c|c|}
			\hline
			Término & Definición\\
			\hline \hline
			\scalebox{0.75}{$A_{Q(11,12,14,21-23,31,41)}$}& $0$ \\ \hline
			\scalebox{0.75}{$A_{Q(13,24)}$} & $\frac{1}{\Delta_Q}$ \\ \hline
			\scalebox{0.75}{$A_{Q32}$} & $\frac{1}{4\Delta_Q}{m_p}^2{L_p}^2L_rg\gamma$ \\ \hline
			\scalebox{0.75}{$A_{Q33}$} & $-\frac{1}{\Delta_Q}\left(f_r\gamma+K_mK_g\right)\left(J_p+\frac{1}{4}m_p{L_p}^2\right)$ \\ \hline
			\scalebox{0.75}{$A_{Q34}$} & $-\frac{1}{2\Delta_Q}m_pL_pL_rf_p\gamma$ \\ \hline
			\scalebox{0.75}{$A_{Q42}$} & $\frac{1}{2\Delta_Q}L_pm_pg\gamma\left(m_r{\left(\frac{L_r}{2}\right)}^2+m_pL_r^2+J_r\right)$ \\ \hline
			\scalebox{0.75}{$A_{Q43}$} & $-\frac{1}{2\Delta_Q}m_pL_pL_r\left(f_r\gamma+K_mK_g\right)$\\ \hline
			\scalebox{0.75}{$A_{Q44}$} & $-\frac{1}{\Delta_Q}f_p\gamma\left(m_r{\left(\frac{L_r}{2}\right)}^2+m_pL_r^2+J_r\right)$ \\ \hline
			\scalebox{0.75}{$B_{Q(11,21)}$} & $0$ \\ \hline
			\scalebox{0.75}{$B_{Q31}$} & $\frac{1}{\Delta_Q}\left(J_p+\frac{1}{4}m_p{L_p}^2\right)$ \\ \hline
			\scalebox{0.75}{$B_{Q41}$} & $\frac{1}{2\Delta_Q}m_pL_pL_r$ \\ \hline
			\multicolumn{2}{|c|}{\scalebox{0.75}{$\Delta_Q= \gamma\left(m_r\left(\frac{L_r}{2}\right)^2+m_pL_r^2+J_r\right)\left( m_p\left(\frac{L_p}{2}\right)^2+J_p\right)-\frac{1}{4}m_p^2L_p^2L_r^2$}}\\ \hline
		\end{tabular}
	\end{center}
\end{table}

Q y R son matrices definidas heurísticamente de acuerdo a una ponderación realizada por el diseñador para las variables de estado del sistema en ambos lazos; estas matrices buscan minimizar 
a partir de la ecuación \eqref{eq5} la energía empleada por las variables de estado ($Q$) y por su entrada ($R$); dando como resultado una  realimentación óptima según la ley de control ($u=-Kx$), en donde la ganancia, al tener un comportamiento continuo en el tiempo, se vuelve una función, que se obtiene a partir de la derivación e igualación a cero de la ecuación \eqref{eq5}, dando como resultado la función de la ecuación \eqref{eq6}.

\begin{equation}
K(t)=-R^{-1}B^TP(t)
\label{eq6}
\end{equation}

Donde en este caso, P es la solución de la ecuación de Riccati \cite{Mohan}, mostrada en la ecuación \eqref{eq7}.

\begin{equation}
A^TP+PA-PBR^{-1}B^TP+Q=0
\label{eq7}
\end{equation}

\subsection{Control por modos deslizantes (SMC) para sistemas en tiempo-discreto}
El control por modos deslizantes (SMC) es un t\'ecnica de dise\~no de controladores, la cual apareci\'o en el contexto de sistemas din\'amicos de estructura variable en la d\'ecada de los 70s, estos sistemas se caracterizan por la presencia de discontinuidades en sus din\'amicas, lo cual hace que sus trayectorias y volumen de fase sean discontinuos en algunos puntos, o usando lenguaje matem\'atico, sean sistemas Lipschitz de forma local. La idea tras los modos deslizantes es aprovechar estas discontinuidades para hacer que las trayectorias alcancen una variedad  en el volumen de fase y se estabilice en esta zona, llevando los estados asint\'oticamente al origen. Las principales ventajas  de esta t\'ecnica se fundamenta en: control y estabilizaci\'on en presencia de incertidumbres acotadas, par\'ametros desconocidos del sistema, o din\'amicas parasitas. El desarrollo de esta t\'ecnica ha sido basada en los trabajos e investigaci\'on de \cite{Shtessel2014} con aplicaciones recientes.

\begin{table}[htbp]\centering	
	\caption{Parámetros matrices de estado plataforma NxtWay}
	\label{table:tabla6}	
	\begin{tabular}{|c|c|}
		\hline
		Término & Definición\\
		\hline \hline
		\scalebox{0.75}{$A_{N(11,12,14,21,22,23,31,41)}$} & $0$ \\ \hline
		\scalebox{0.75}{$A_{N(13,24)}$} & $1$ \\ \hline
		\scalebox{0.75}{$A_{N32}$} & $-\frac{gML\left(MLR-2n^2J_m\right)}{\Delta_N}$ \\ \hline
		3{*}{\scalebox{0.75}{$A_{N33}$}} 
		&\scalebox{0.75}{ $-\frac{1}{\Delta_N}2\left(\beta{}+f_w\right)$}\\
		&\scalebox{0.75}{$\left(ML^2+J_{q_2}+2n^2J_m\right)+$}\\
		&\scalebox{0.75}{$\beta{}\left(MLR-2n^2J_m\right)$} \\ \hline
		\scalebox{0.75}{$A_{N34}$} & $\frac{2\beta{}\left(ML^2+J_{q_2}+MLR\right)}{\Delta_N}$ \\ \hline
		\scalebox{0.75}{$A_{N42}$} & $\frac{MgL\left(2mR^2+MR^2+{2J}_w+{2n}^2J_m\right)}{\Delta_N}$ \\ \hline
		3{*}{\scalebox{0.75}{$A_{N43}$}} & \scalebox{0.75}{$\frac{1}{\Delta_N}2\left(\beta{}+f_w\right)(MLR-2n^2J_m+$}\\
		&\scalebox{0.75}{$\beta{}(2mR^2+MR^2+2{J}_w+$}\\
		&\scalebox{0.75}{${2n}^2J_m))$} \\ \hline
		\scalebox{0.75}{$A_{N44}$} & $-\frac{2\beta{}\left(MLR+2mR^2+MR^2+2{J}_w\right)}{\Delta_N}$ \\ \hline
		\scalebox{0.75}{$B_{N(11,12,21,22)}$} & $0$ \\ \hline
		\scalebox{0.75}{$B_{N(31,32)}$} & $\frac{\alpha{}\left(ML^2+J_{q_2}+MLR\right)}{\Delta_N}$ \\ \hline
		\scalebox{0.75}{$B_{N(41,42)}$} & $-\frac{\alpha{}\left(MLR+2mR^2+MR^2+2{J}_w\right)}{\Delta_N}$ \\ \hline
		\multicolumn{2}{|c|}{\scalebox{0.75}{$\Delta_N=\left(\left(2m+M\right)R^2+2J_W+2\eta^2J_m\right)\left(ML^2+J_{q_2}+2\eta^2J_m\right)-\left(MLR-2\eta^2J_m \right)^2$}}\\ \hline		
	\end{tabular}
\end{table}

Para el diseño se trabaja con un equivalente discreto del sistema original basado en una t\'ecnica de muestreo y retenci\'on aplicando el retenedor de orden cero (ZOH), un espacio de estados discreto es establecido. De forma m\'as simple, la aproximaci\'on es, desde el punto de vista del controlador, la forma como 'observa' el sistema din\'amico. Para el dise\~no del controlador es importante definir dos momentos en la metodolog\'ia, el primero consiste en la definici\'on adecuada de una variedad estable, luego, se selecciona una ley de control en lazo cerrado que incluye una ganancia de realimentaci\'on y una funci\'on discontinua definida a partir de un an\'alisis basado en el teorema de Lyapunov.

Se tiene una representaci\'on en espacio de estados definida como:
\begin{equation}
\begin{array}{c}
x_{1(k\text{+1)}}=A_{11}x_{1(k\text{)}}+A_{12}x_{2(k\text{)}}\\
x_{2(k\text{+1)}}=A_{21}x_{1(k\text{)}}+A_{22}x_{2(k\text{)}}+u.
\end{array}
\label{equa1}
\end{equation}
Donde $x_{1(k)}\in\mathbb{R}^{n-m}$, $x_{2(k)}\in\mathbb{R}^{m}$,$A_{i,j}$ es una matriz real de  par\'ametros constantes.

Entonces se define la variedad estable como:

\begin{equation}
s_{k}=Cx_{1(k\text{)}}+x_{2(k)}=0\Rightarrow x_{2(k)}=-Cx_{1(k\text{)}}.
\label{equa2}
\end{equation}
Donde $C\in\mathbb{R}^{m\times(n-m)}$.

Reemplazando esta ley de control en el sistema original del bloque superior, se obtiene:
\begin{equation}
x_{1(k\text{+1)}}=\left(A_{11}-A_{12}C\right)x_{1(k\text{)}}.
\label{equa3}
\end{equation}
Haciendo una selecci\'on adecuada de los valores de la matriz $C$ se pueden ajustar los autovalores de la matriz resultante $A_{11}-A_{12}C$ dentro del    c\'irculo unitario.

Una vez se ha considerado la selecci\'on de la variedad estable entonces se procede a determinar la ley de control por modos deslizantes como: 
\begin{equation}
u_{k}=-ksign(s_{k}).
\label{equa4}
\end{equation}
Donde $sign$ es la funci\'on signo y $k$ es un valor a ser determinado basado en el an\'alisis de estabilidad aplicando el segundo m\'etodo de Lyapunov. Para ello sea $V\::\:\mathbb{R}^{k}\rightarrow\mathbb{R}$ una funci\'on que cumple con $V(0)=0$ y $V>0$ estrictamente mayor a $0$. Su primera variación a lo largo de las trayectorias esta dada por $\Delta V(x_{k})=V(x_{k+1})-V(x_{k})$ la cual cumple con $\Delta V(x_{k})\leq-T_{s}\alpha V(x_{k})^{\frac{1}{2}}$  $\alpha$ siendo real positivo y $T_{s}$ el tiempo de muestreo sea suficientemente pequeño, lo cual garantiza que el sistema alcance estabilidad en tiempo finito.
Considerando la siguiente funci\'on de Lyapunov 
\begin{equation}
V=\frac{1}{2}s_{k}^{2}.
\end{equation}
Aplicando la primera variaci\'on
\begin{equation}
\Delta V=\frac{1}{2}\left(s_{k+1}^{2}-s_{k}^{2}\right).
\end{equation}
Para aplicar esta funci\'on de Lyapunov se tiene:
\begin{equation}
s_{k}=\alpha_{1}x_{1(k)}+\alpha_{2}x_{1(k)}+\alpha_{3}x_{3(k)}+x_{4(k)}=0.
\end{equation}
En forma compacta $s_{k}=\mathbf{L}\mathbf{x}_{k}$, donde $\mathbf{L}\in\mathbb{R}^{1\times n}$ y $\mathbf{x}_{k}$ es el vector de estado. Aplicando el desplazamiento en $k+1$
\begin{equation}
s_{k+1}=\mathbf{K}\mathbf{x}+u.
\end{equation}
Donde $\mathbf{K}$ es un combinaci\'on de par\'ametros entre la variedad definida  y la matriz de estados del sistema original. 
Juntando t\'erminos se obtiene:
\begin{equation}
\frac{1}{2}\left(\mathbf{x^{T}K^{T}Kx}+2(\mathbf{Kx}+u_{k})+u_{k}^{2}-\mathbf{x_{k}^{T}L^{T}Lx_{k}}\right)> 0.
\end{equation}
Considerando en este proceso que $s_{k}=0$, se puede inferir adem\'as que $s_{k+1}=0$, y sustituyendo por la ley de control se tiene: 
\begin{equation}
u_{k}=-\mathbf{Kx}-\mathit{k}sign(s_{k}).
\label{equa5}
\end{equation}
Simplificando la ecuaci\'on y encontrando una cota para la selecci\'on de $k$:
\begin{equation}
\begin{array}{c}
\frac{1}{\sqrt{2}}\left(k^{2}-1\right)^{\frac{1}{2}}\left|s_{k}\right|\leq-T_{s}\frac{\alpha}{\sqrt{2}}\left|s_{k}\right|\\
k\leq\left(\left(-T_{s}\frac{\alpha}{\sqrt{2}}\right)^{2}+1\right)^{\frac{1}{2}}\:\alpha>0
\end{array}.
\end{equation}
Donde $T_{s}$ es el tiempo de muestreo, con esta condici\'on se garantiza en el sistema su estabilidad en tiempo finito.

\section{Implementaci\'on}

El proceso de implementación inicia a partir de las técnicas diseñadas, tomando en cuenta las restricciones hardware y software de las plataformas abordadas. 

Las restricciones software de cada plataforma se relacionan con los entornos de desarrollo. La plataforma Quanser utiliza el entorno de programación gráfico de Matlab, haciendo uso del Quarc Toolbox; mientras que la plataforma NxtWay utiliza el entorno de desarrollo RobotC de Mindstorms. Así, una característica propia en el ambiente de desarrollo Matlab, es la posibilidad de implementar algoritmos en tiempo continuo, donde el proceso de discretización se hace transparente en el desarrollo de aplicaciones que contienen modelos en Simulink. Caso contrario, ocurre en RobotC, donde la ley de control que se implementa debe ser discreta, dado que la programación se realiza a bajo nivel en lenguaje C. Otra restricción de software se encuentra en relación al muestreo de las variables de estado, para el caso de RobotC, la única manera de exportar datos del bloque controlador NXT, es a través de su consola de texto, que se sincroniza con los procesos internos del bloque cada 10ms como máximo, tiempo que no siempre coincide con el tiempo de aplicación. En Matlab por su parte, el muestro de las variables de estado, si se puede efectuar en tiempo de aplicación, pero existe un limite de datos que se pueden capturar, dependiendo de los recursos físicos como memoria, que se reservan para almacenar datos dentro de los indicadores gráficos de Simulink.

Por otro lado, en cuanto a las restricciones hardware donde se implementaron los algoritmos, se tiene que las frecuencias de procesamiento de las estructuras son diferentes, la frecuencia del bloque controlador Quanser, limitada por su tarjeta de adquisición de datos, es el doble de la frecuencia de bloque NxtWay, por lo que la cantidad de datos capturados deben ser diferentes para realizar comparaciones en un tiempo de simulación igual.

\subsection{Control LQR}

Para el controlador  LQR se usan los siguientes valores en el diseño: Las matrices $Q_x$ y $R_x$ empleadas en ambas plataformas para el cálculo del controlador son:

\begin{equation}
\footnotesize
Q_Q=\left[\begin{array}{
	cccc}
5 & 0 & 0 & 0 \\
0 & 1 & 0 & 0 \\
0 & 0 & 1 & 0 \\
0 & 0 & 0 & 1
\end{array}\right]\quad R_Q=1
\label{eq8}
\end{equation}

\begin{equation}
\footnotesize
Q_N=\left[\begin{array}{
	ccccc}
1 & 0 & 0 & 0 & 0 \\
0 & 6x10^5 & 0 & 0 & 0 \\
0 & 0 & 1 & 0 & 0 \\
0 & 0 & 0 & 1 & 0 \\
0 & 0 & 0 & 0 & 4x10^2
\end{array}\right]
\quad
R_N=\begin{bmatrix}
1x10^3 & 0 \\
0 & 1x10^3 
\end{bmatrix}
\label{eq9}
\end{equation}

\begin{figure}[htbp]
	\centering
	\includegraphics[width=9cm]{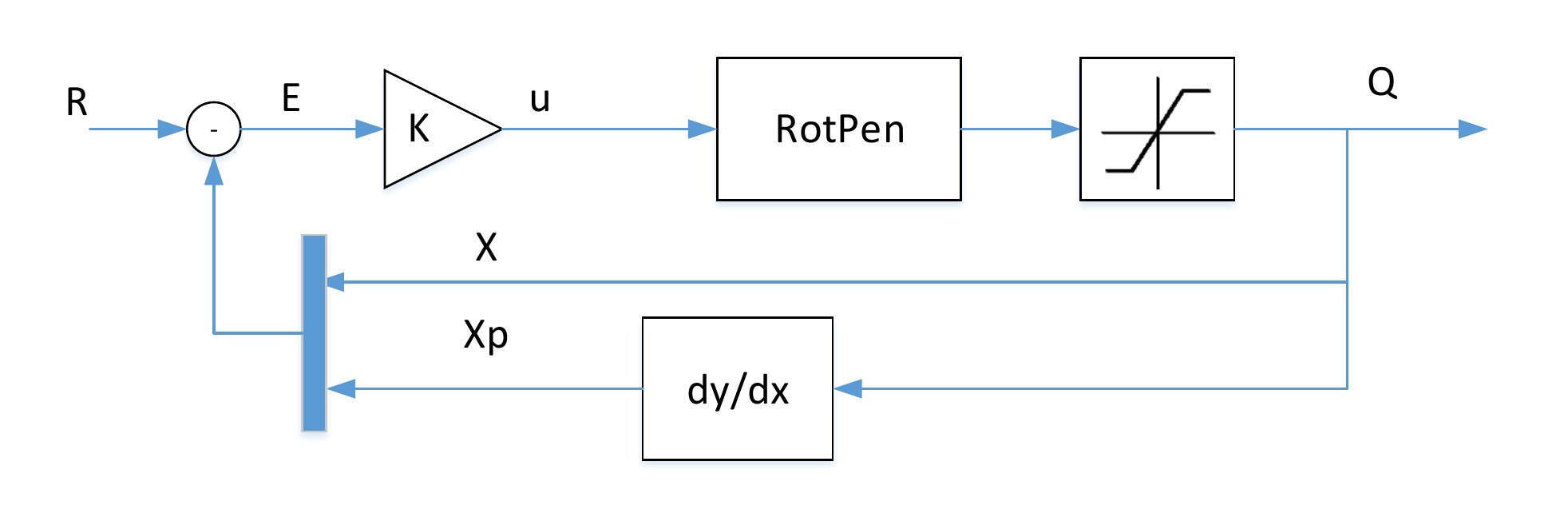}
	\caption{Diagrama de bloques control LQR Péndulo RotPen}
	\label{fig:bloquesQ}
\end{figure}

\begin{figure}[htbp]
	\includegraphics[width=9cm]{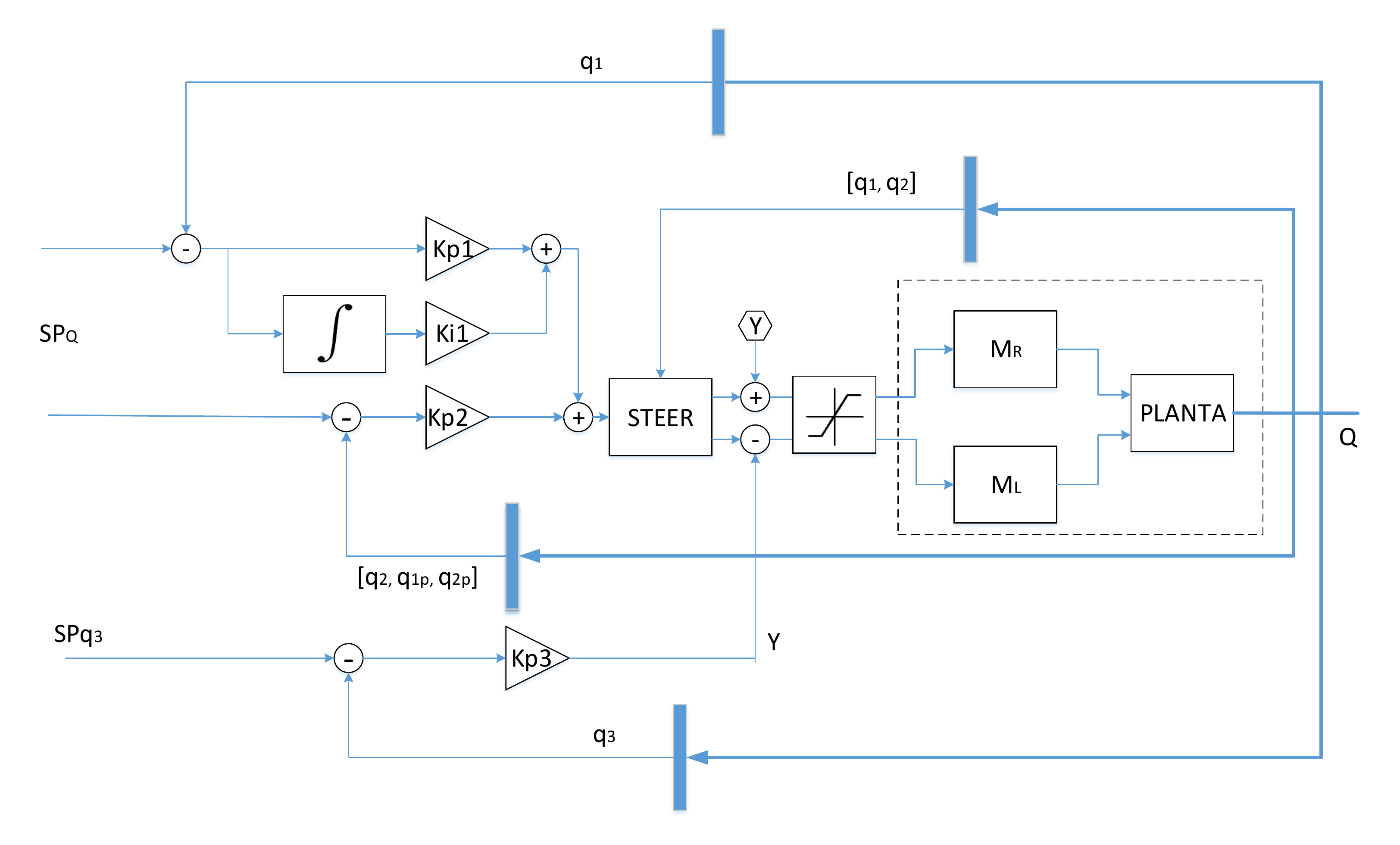}
	\caption{Diagrama de bloques control LQR Péndulo NxtWay}
	\label{fig:bloquesN}
\end{figure}

Se observa en la ecuación \ref{eq9}, que para el caso del NxtWay, se incluye una columna en los valores de Q y R, por el primer lazo de control utilizado, en donde se adiciona una quinta variable de estado definida por la integral de $q_1$ para garantizar la posición de la plataforma dada su independencia del modelo dinámico. Los lazos de control se presentan en los diagramas de bloques de las figuras \ref{fig:bloquesQ} y \ref{fig:bloquesN}. La figura \ref{fig:bloquesQ} muestra el diagrama de bloques de la plataforma RotPen, que toma la forma de un sistema de control realimentado LQR; por otro lado, la figura \ref{fig:bloquesN} expone el diagrama de bloques de la plataforma NxtWay; adicionalmente, se realiza una corrección de la posición del móvil debido a los deslizamientos y asincronismos en el movimiento que puedan ocurrir cuando la energía de actuación de la rueda virtual se deriva para cada uno de los dos actuadores físicos, este bloque se denomina Steer. Así mismo, se incluye un bloque de saturación sobre la señal de actuación, con un valor igual al máximo voltaje en operación normal, para los motores de cada planta. Los valores se observan en las Tablas \ref{table:tabla1} y \ref{table:tabla2}.

De la deducción de las ecuaciones \eqref{eq5} y \eqref{eq7} se obtienen las constantes de realimentación de control de los sistemas, estas se observan en la ecuación \eqref{eq10} para Quanser y \eqref{eq11} para NxtWay.

\begin{equation}
K_Q=\left[-2.2361\ 25.4512-2.4613\ 3.6332\right]\
\label{eq10}
\end{equation}
\begin{equation}
\begin{split}
&K_N=[-0.8211\ -69.4743\ -1.0739\ -9.0738] \\
&K_{Ni}=-0.4472 \\
\end{split}
\label{eq11}
\end{equation}

\begin{figure}[htbp]
	\centering
	\subfigure[QuanserThetaPos]{
		\label{f:Quansert}
		\includegraphics[width=0.23\textwidth]{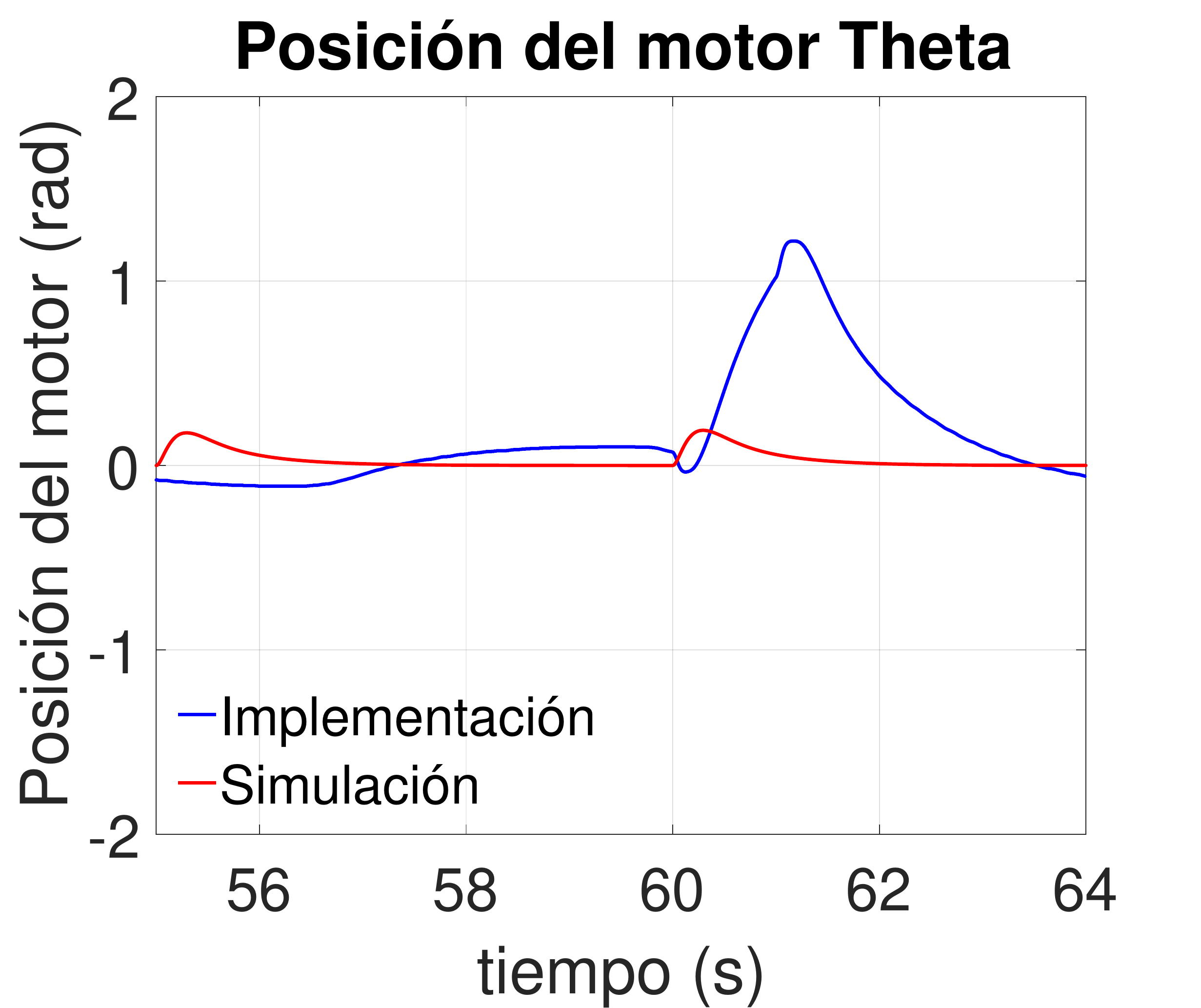}}
	\subfigure[QuanserThetaVel]{
		\label{f:Quansertp}
		\includegraphics[width=0.23\textwidth]{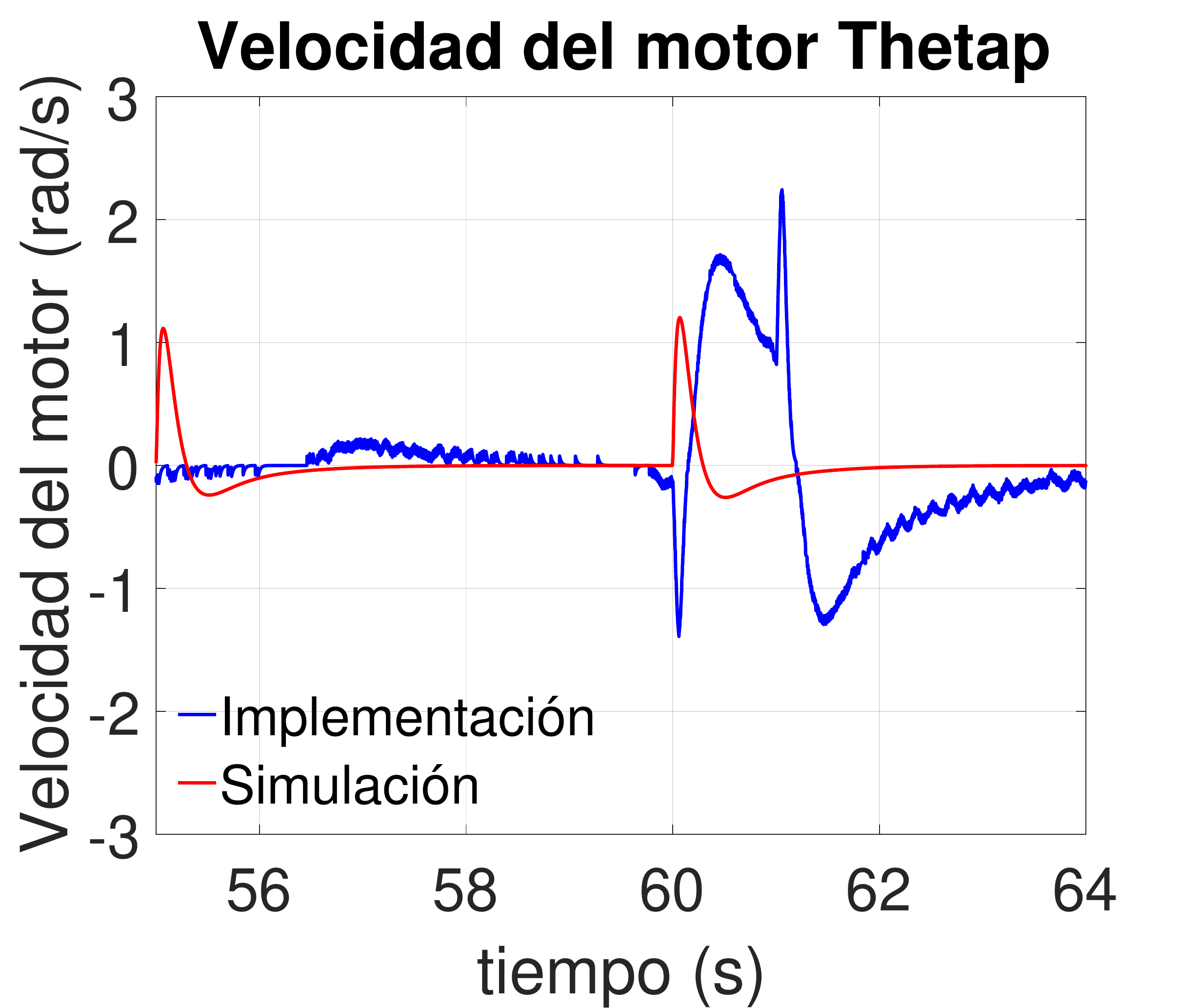}}
	\subfigure[QuanserPsi]{
		\label{f:Quansera}
		\includegraphics[width=0.23\textwidth]{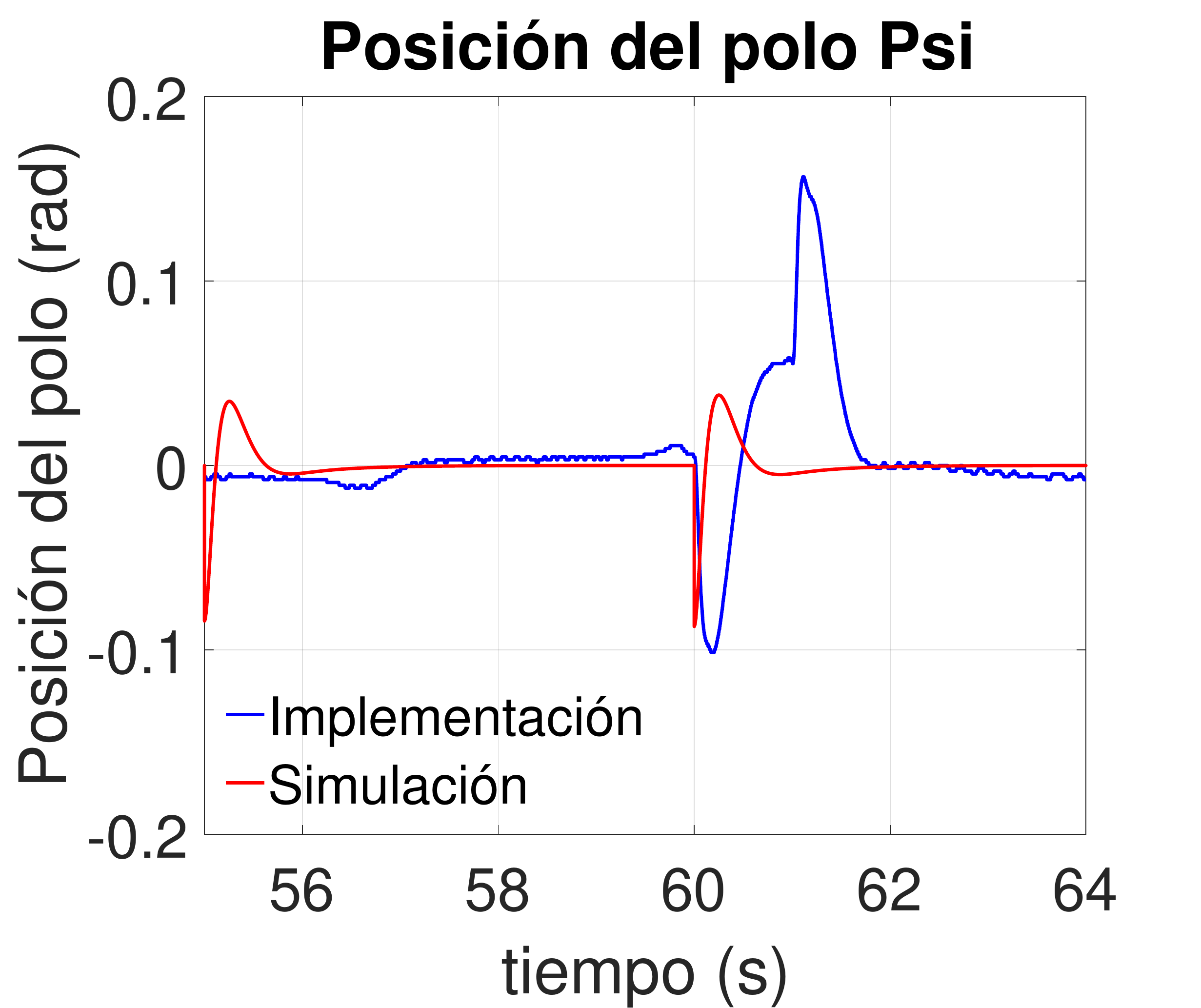}}
	\subfigure[QuanserPsiVel]{
		\label{f:Quanserap}
		\includegraphics[width=0.23\textwidth]{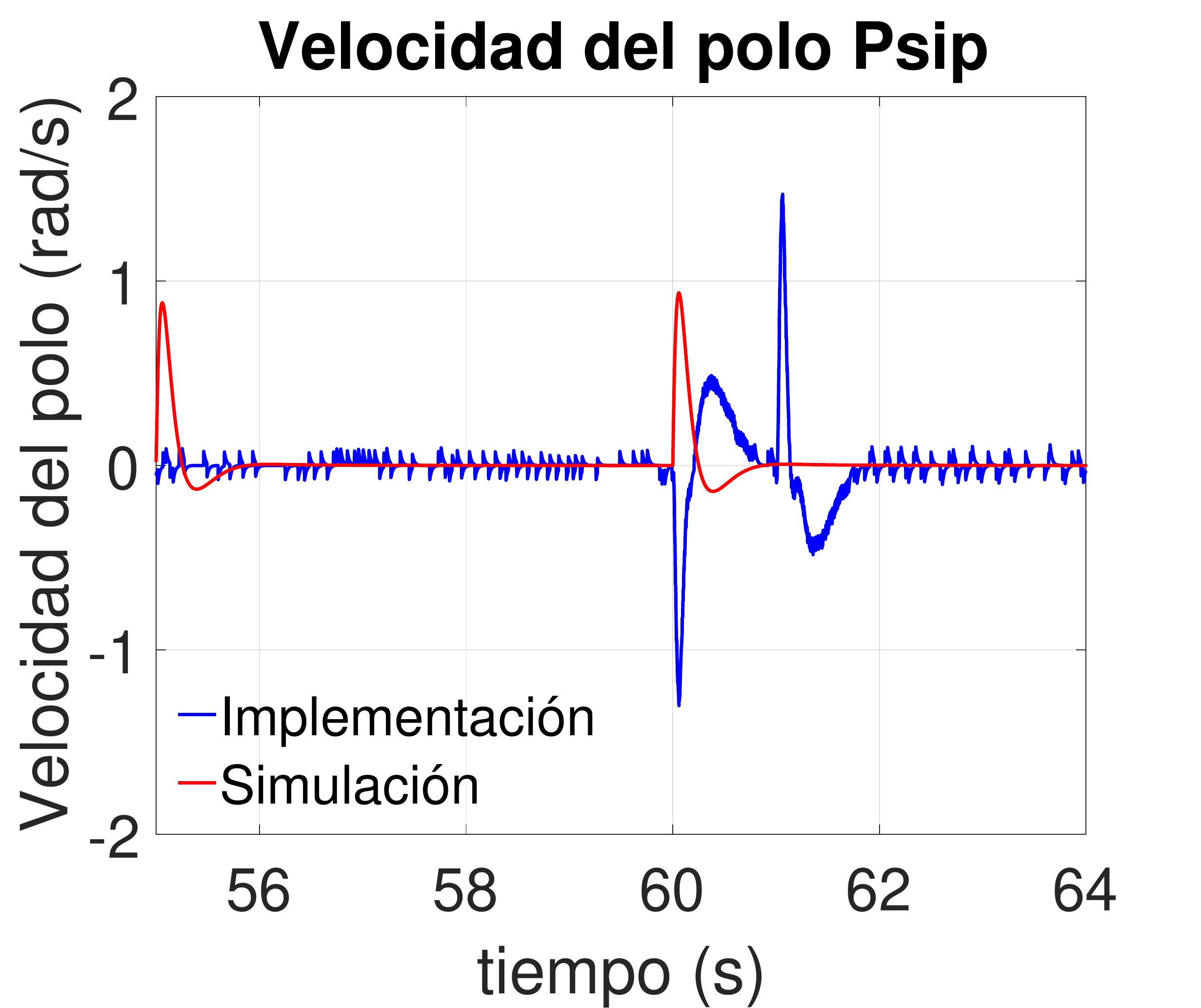}}
	\subfigure[QuanserSalida]{
		\label{f:QuanserU}
		\includegraphics[width=0.23\textwidth]{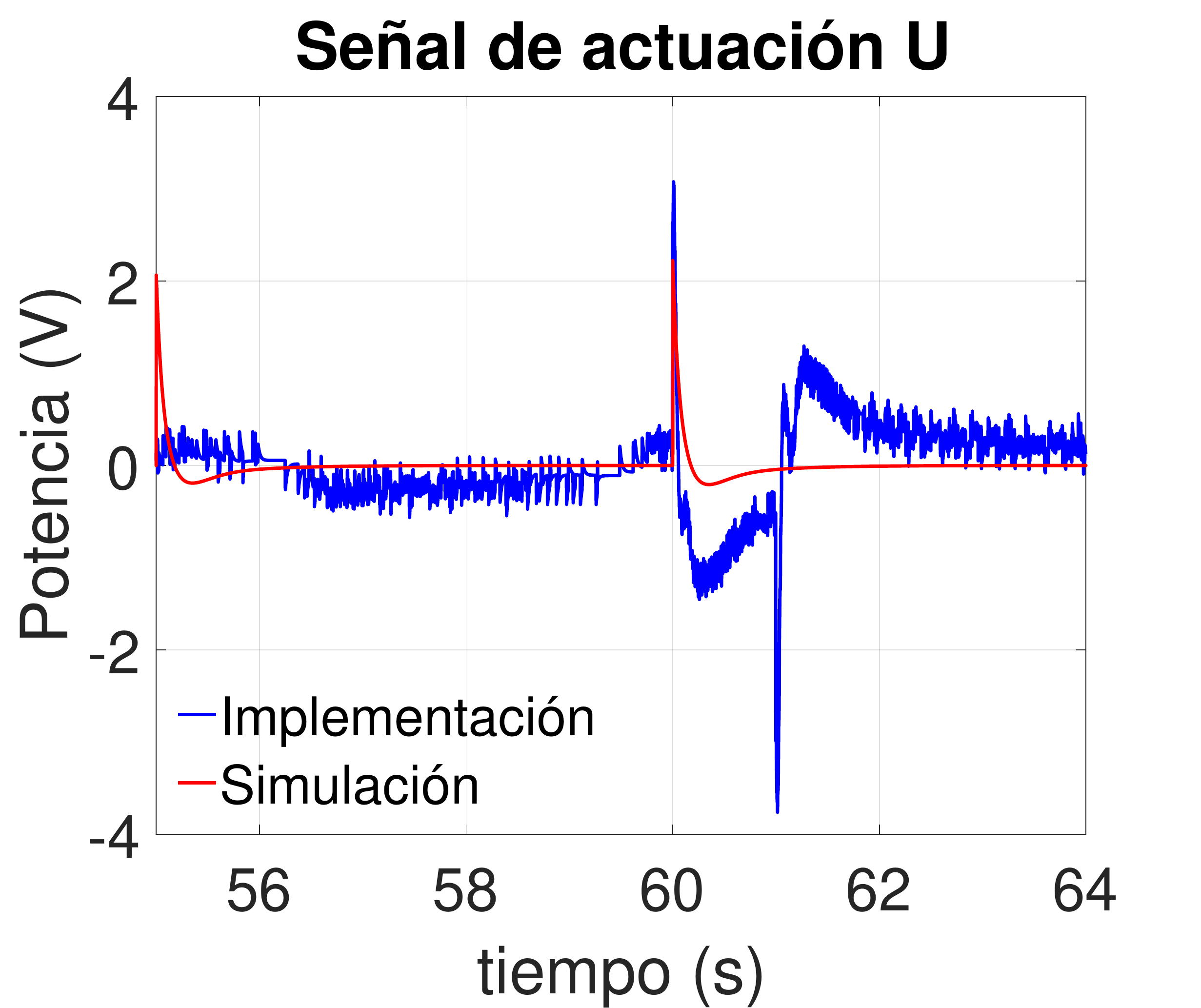}}
	\caption{Respuesta control LQR Péndulo RotPen}
	\label{fig:fig5}
\end{figure}

El comportamiento dinámico del modelo RotPen simulado e implementado se presenta en la figura \ref{fig:fig5} ante una perturbación, que inicia después de 60 segundos, la cual consiste en añadir un voltaje extra al actuador, proporcional al voltaje máximo permitido de cada plataforma. Para cada sistema, el punto de referencia se encuentra en 0 (Independientemente de las unidades), en la figura \ref{fig:fig5} se observa que el controlador estabiliza cada una de las variables de estado en su referencia, en un tiempo de 2 a 3 segundos aproximadamente; es decir, en el rango de 62 a 63 segundos como se muestra en las figuras. Asimismo, presenta un sobreimpulso menor con el modelo simulado que con el modelo implementado y una señal de actuación menor a 3 voltios en escala de 0 a 10 v.

\begin{figure}[htbp]
	\centering
	\subfigure[NxtWayThetaPos]{
		\label{f:Nxtwayt}
		\includegraphics[width=0.23\textwidth]{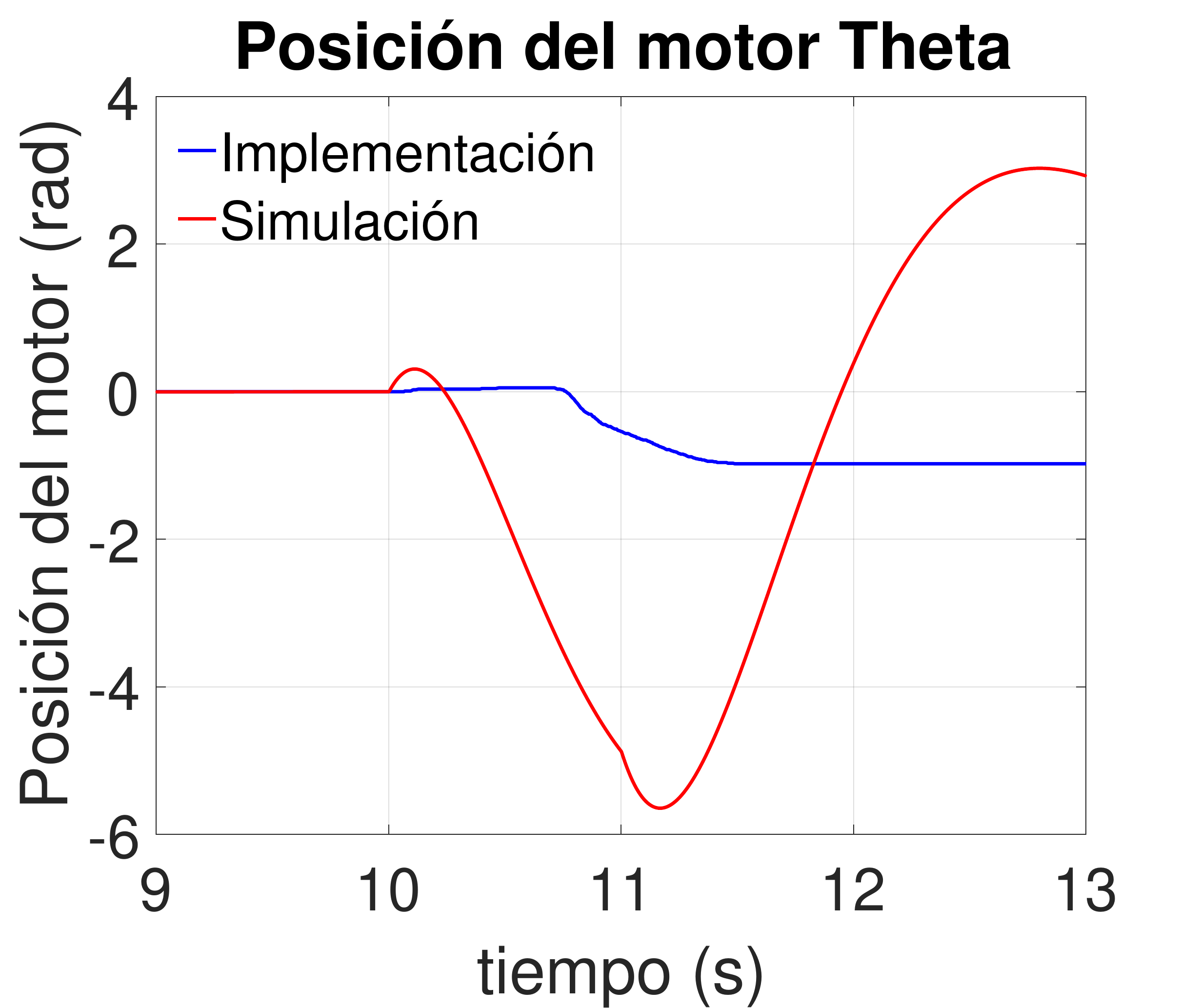}}
	\subfigure[NxtWayThetaVel]{
		\label{f:Nxtwaytp}
		\includegraphics[width=0.23\textwidth]{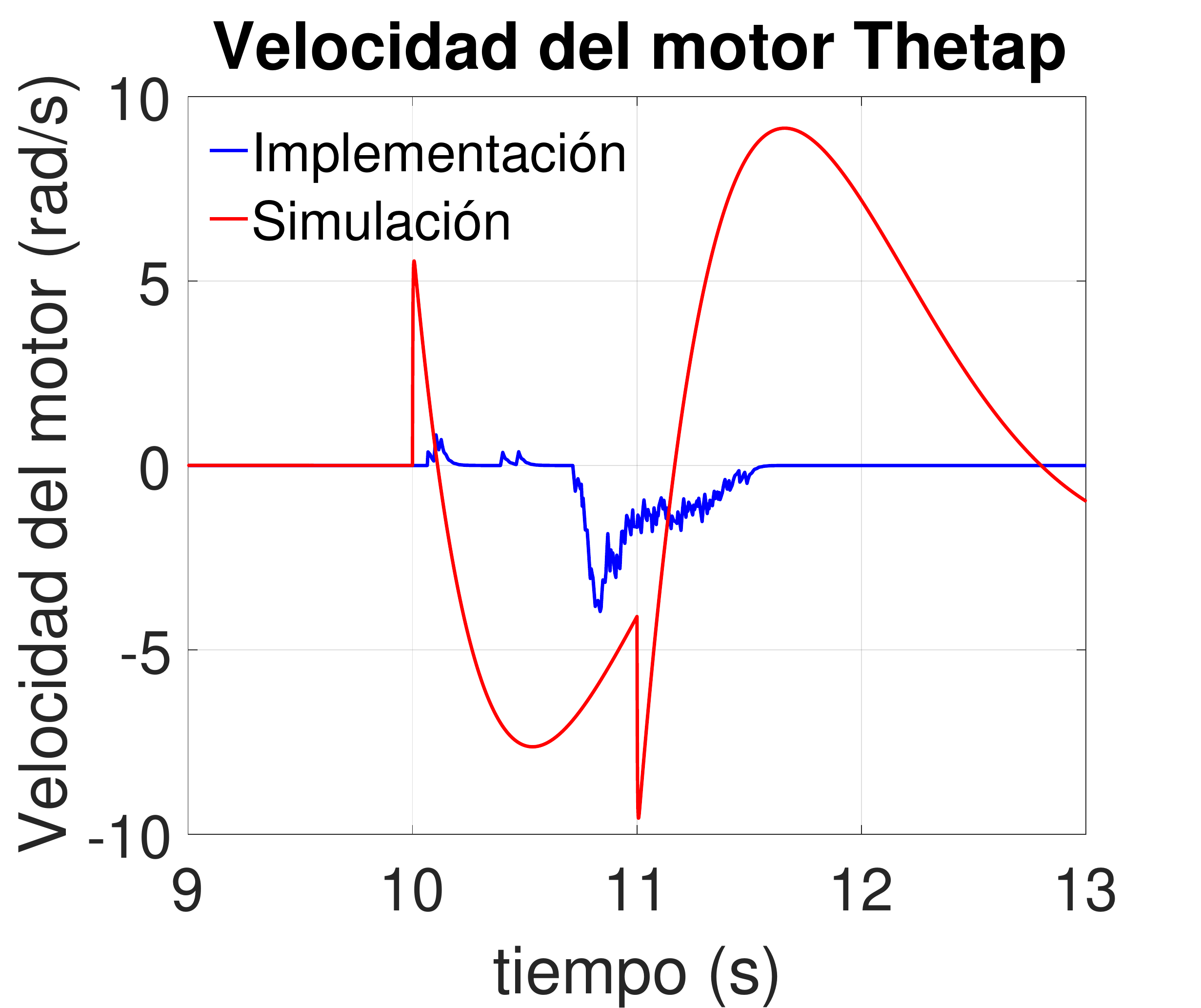}}
	\subfigure[NxtWayPsi]{
		\label{f:nxtwaya}
		\includegraphics[width=0.23\textwidth]{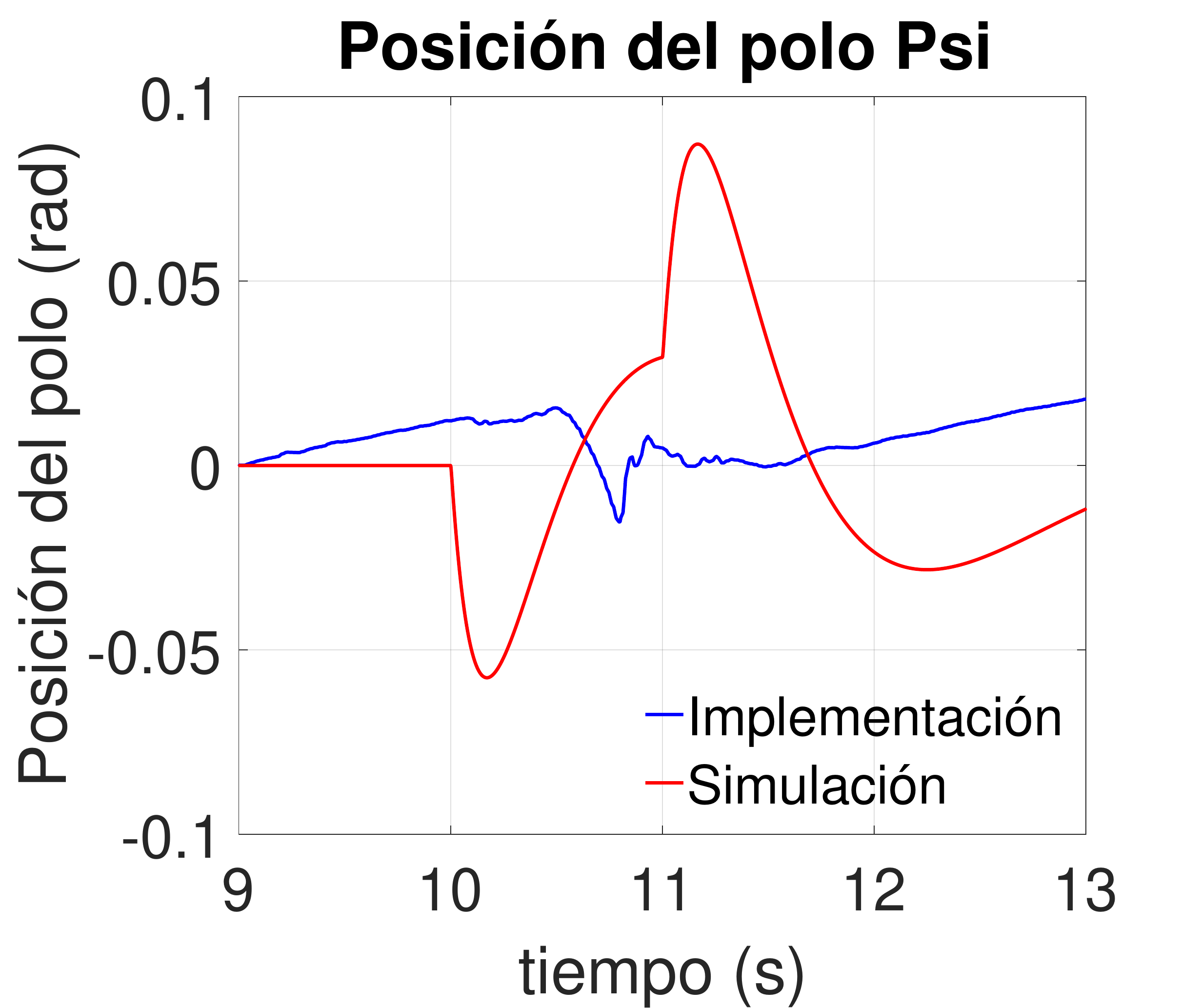}}
	\subfigure[NxtWayPsiVel]{
		\label{f:nxtwayap}
		\includegraphics[width=0.23\textwidth]{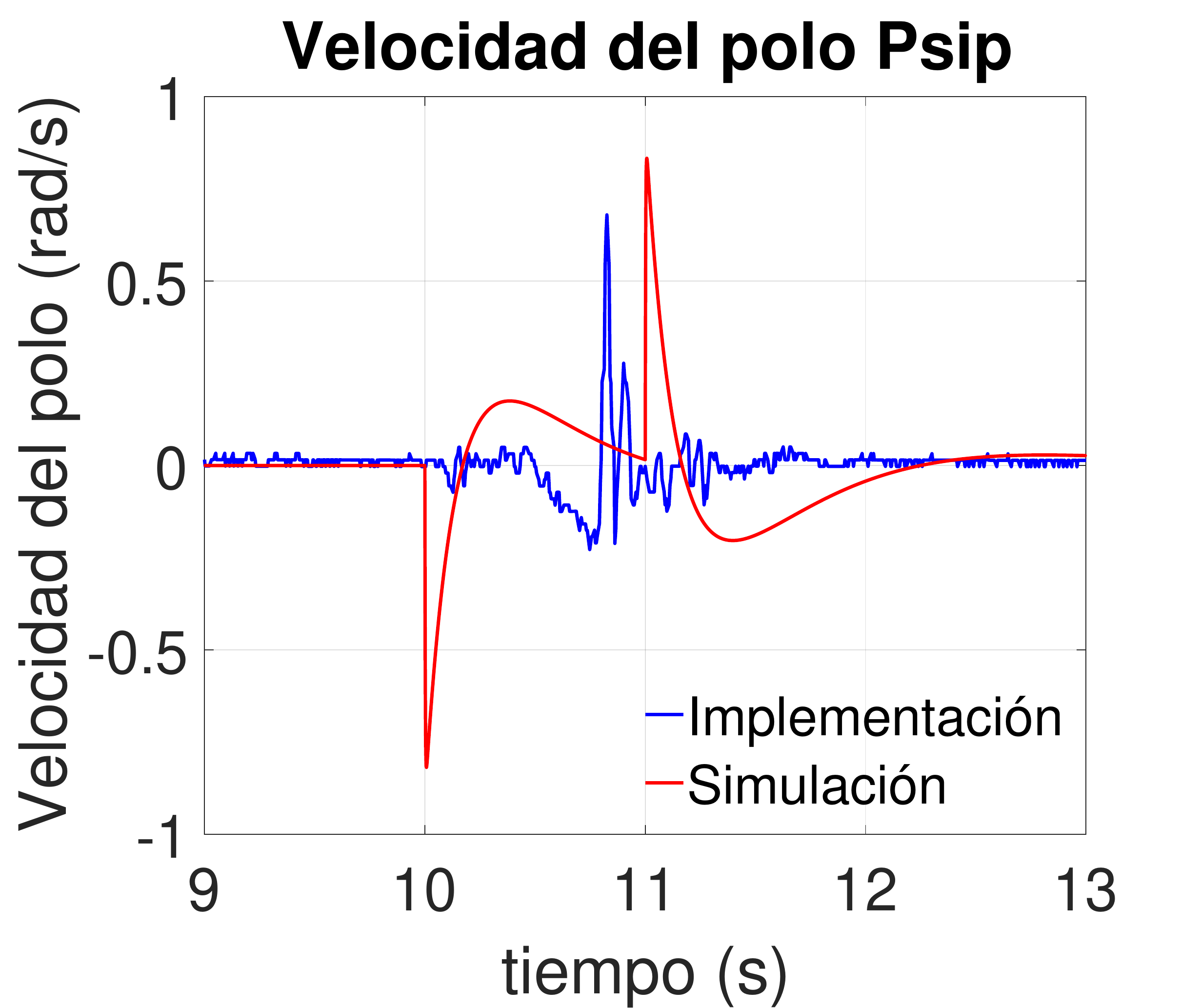}}
	\subfigure[NxtWaySalida]{
		\label{f:NxtwayU}
		\includegraphics[width=0.23\textwidth]{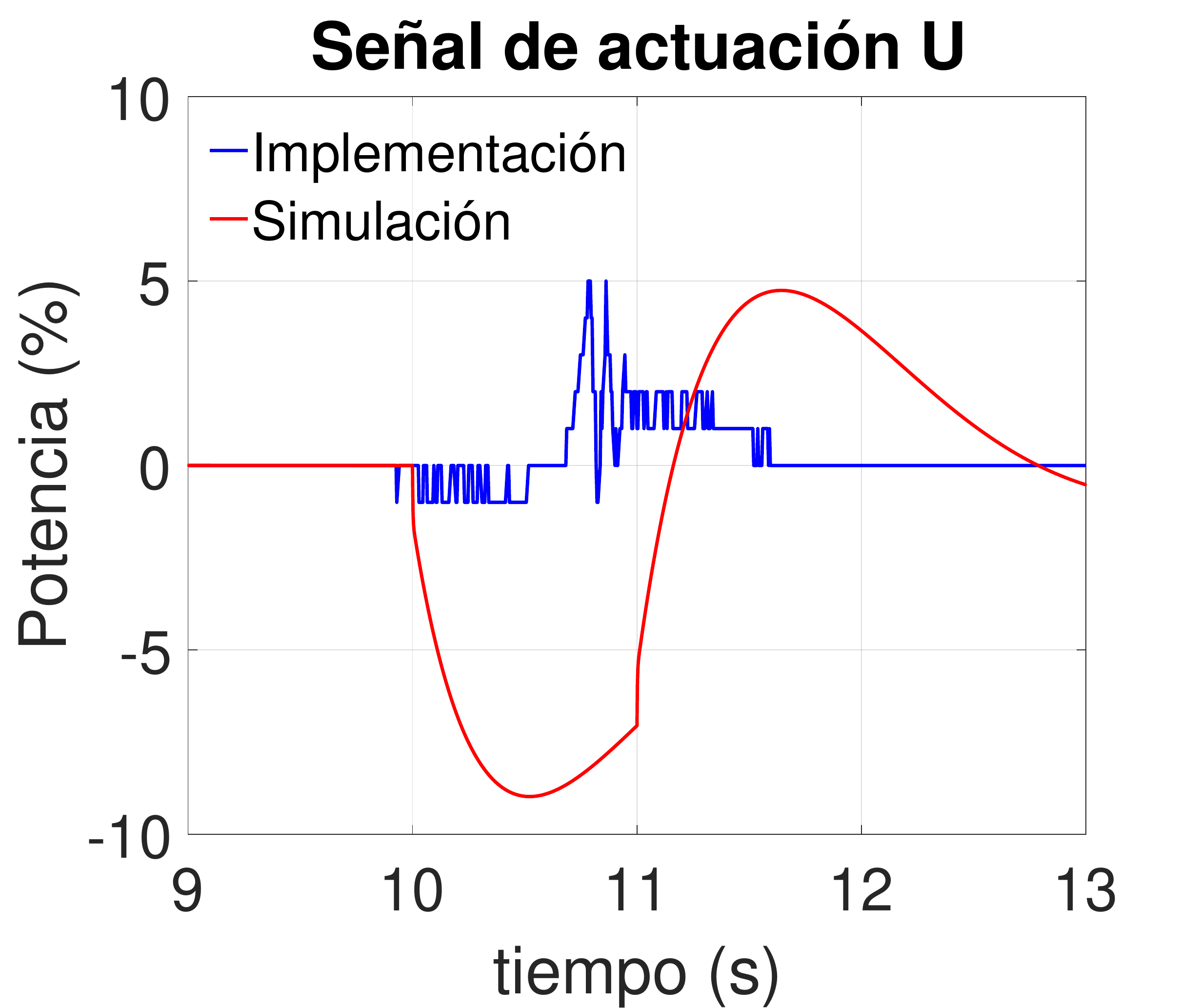}}
	\caption{Respuesta control LQR Péndulo NxtWay}
	\label{fig:fig6}
\end{figure}

De igual manera, se observa en la figura \ref{fig:fig6}, la respuesta de la implementación en el péndulo móvil NxtWay, al igual que en el RotPen, la referencia de todas las variables de estado se encuentra en 0 unidades, a diferencia de la respuesta de la planta de Quanser, las variables de estado poseen un grado de oscilación, en especial aquellas relacionadas con las ruedas, al tener una dinámica independiente del polo.

\subsection{Control por modos deslizantes (SMC)}

Para la implementación de la técnica por modos deslizantes se consideran los siguiente parámetros de diseño para ambas plataformas. El vector $L_{N}$ y $L_{Q}$ est\'an asociados a las variedades estable consideradas en la técnica Estos valores deben garantizar que los autovalores se encuentre en el c\'irculo unitario, una vez estos valores son obtenidos se considera la expresi\'on \ref{equa5}, para seleccionar el vector de realimentaci\'on $K_{N}$ y $L_{Q}$, el cual garantice la desigualdad planteada.

\begin{equation}
\mathbf{K_N}=\left[0\ 1.0194 -3.3149\ 3.2957\right]\
\label{KMayus_SMC}
\end{equation}
\begin{equation}
k_N=20
\label{KMinus_SMC}
\end{equation}
\begin{equation}
\mathbf{L_N}=\left[0\ -0.0002\ 1\right]\
\label{L_SMC}
\end{equation}

\begin{equation}
\mathbf{K_Q}=\left[-4.8606\ 5.3882 -0.334\ 0.3472\right]\
\label{KMayusN_SMC}
\end{equation}
\begin{equation}
k_Q=2.5
\label{KMinusN_SMC}
\end{equation}
\begin{equation}
\mathbf{L_Q}=\left[1155.5\ -4.1 61\ -68.3\right]\
\label{LN_SMC}
\end{equation}

La respuesta de la implementación en el péndulo móvil NxtWay, al igual que en los casos anteriores, plantea una referencia de todas las variables de estado en 0 unidades, a diferencia de la respuesta del control LQR, esta respuesta maneja una oscilación en su señal de actuación provocada por el uso de una ley de control discontinua o con scattering (Conmutación de las dinámicas de control en las dos fases SMC) en el cálculo de la salida del controlador, con los valores de las ecuaciones \eqref{KMayus_SMC}, \eqref{KMinus_SMC} y \eqref{L_SMC}, se obtienen los resultados de las variables de estado y señal de actuación en la figura \ref{fig:smcn}.

De igual manera, la respuesta de implementación del péndulo rotatorio RotPen, plantea un experimento con la referencia de nuevo en 0 unidades, esta respuesta se observa en la figura \ref{fig:smcq}, al igual que el caso del NxtWay maneja un nivel de Scattering en sus señales, provocada por la implementación propia del control, sin embargo al contar Quanser con una plataforma robusta para el procesamiento y discretización de las señales incluidas, su oscilación no maneja valores tan altos como el NxtWay.

Como datos adicionales a discutir en el proceso de implementación, se destaca el uso de filtros derivadores en cada uno de los entornos, desarrollados a partir de la respuesta en alta frecuencia de los giroscopios y encoders de las plataformas, filtrado necesario debido al proceso de discretización de los algoritmos de lectura de las variables, adicional a la respuesta térmica de los elementos y al ruido electromagnético presente en el sistema; de igual manera, se destaca el manejo del tiempo de muestreo de cada algoritmo, que esta limitado por los entornos y es importante para las operaciones realizadas, como el cálculo de las integrales o de las derivadas de las variables obtenidas.

\section{Análisis de Resultados}
\begin{figure}[htbp]
	\centering
	\subfigure[NxtWayThetaPos SMC]{
		\label{f:Nxtwaytsmc}
		\includegraphics[width=0.23\textwidth]{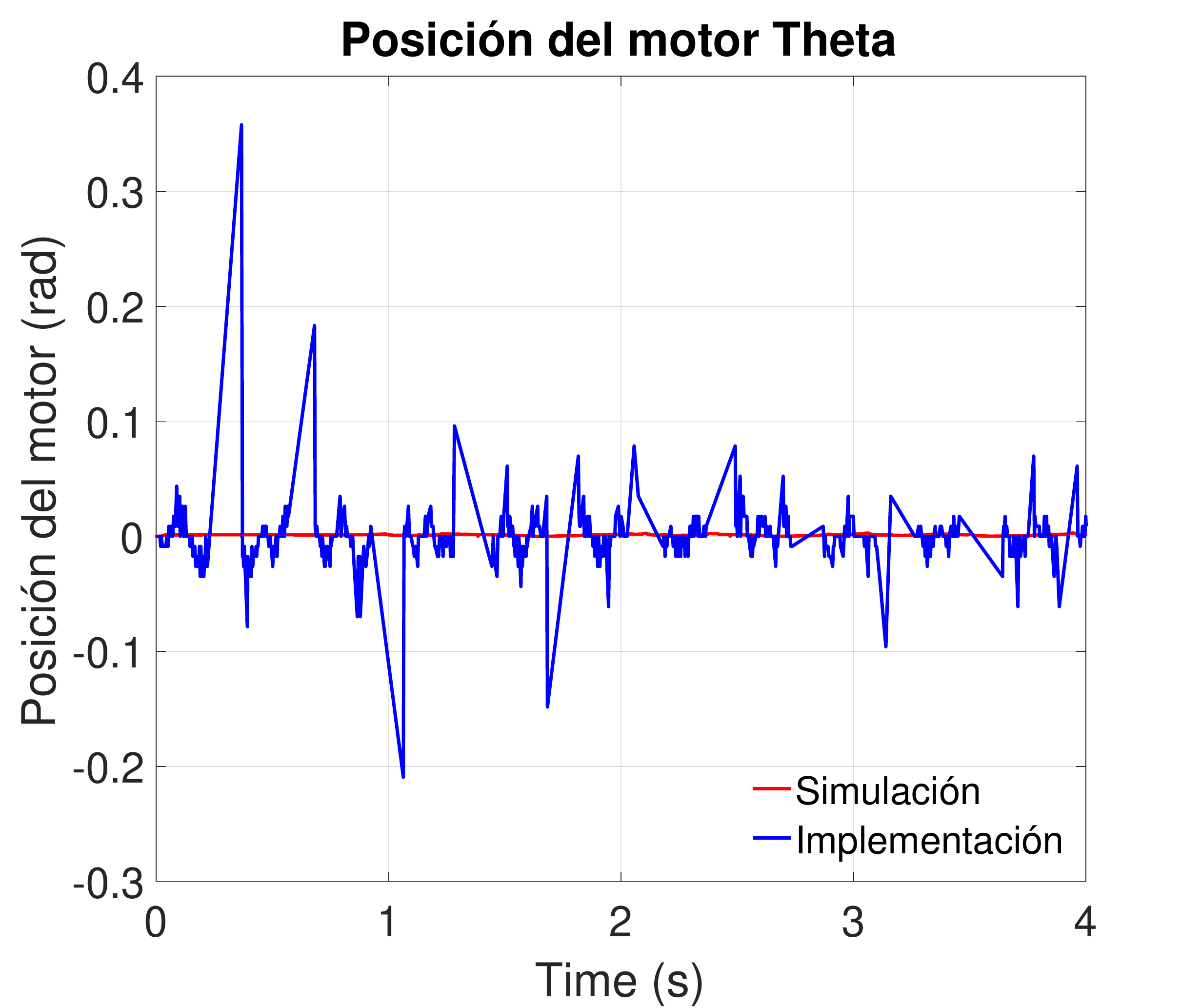}}
	\subfigure[NxtWayThetaVel SMC]{
		\label{f:Nxtwaytpsmc}
		\includegraphics[width=0.23\textwidth]{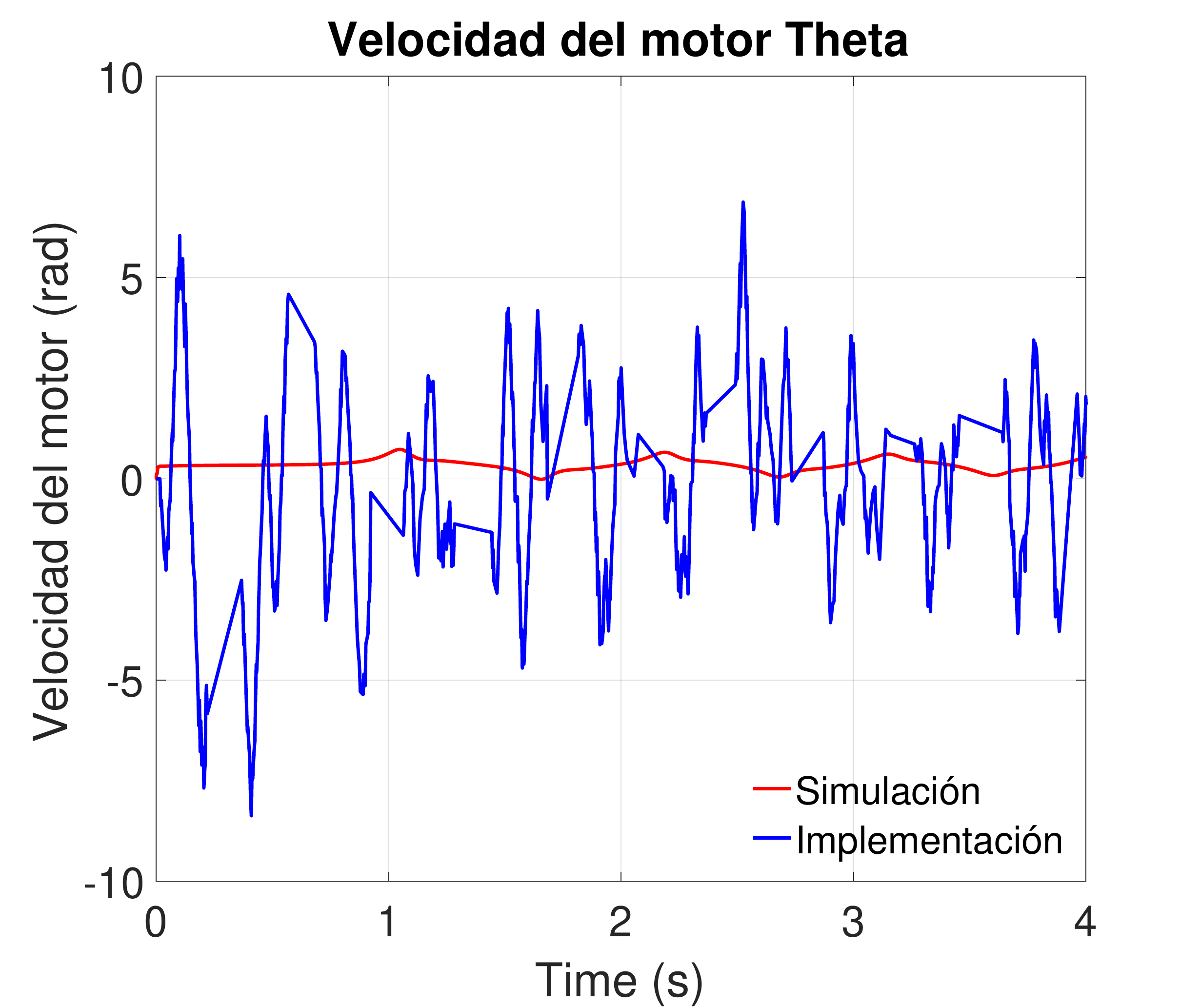}}
	\subfigure[NxtWayPsi SMC]{
		\label{f:nxtwayasmc}
		\includegraphics[width=0.23\textwidth]{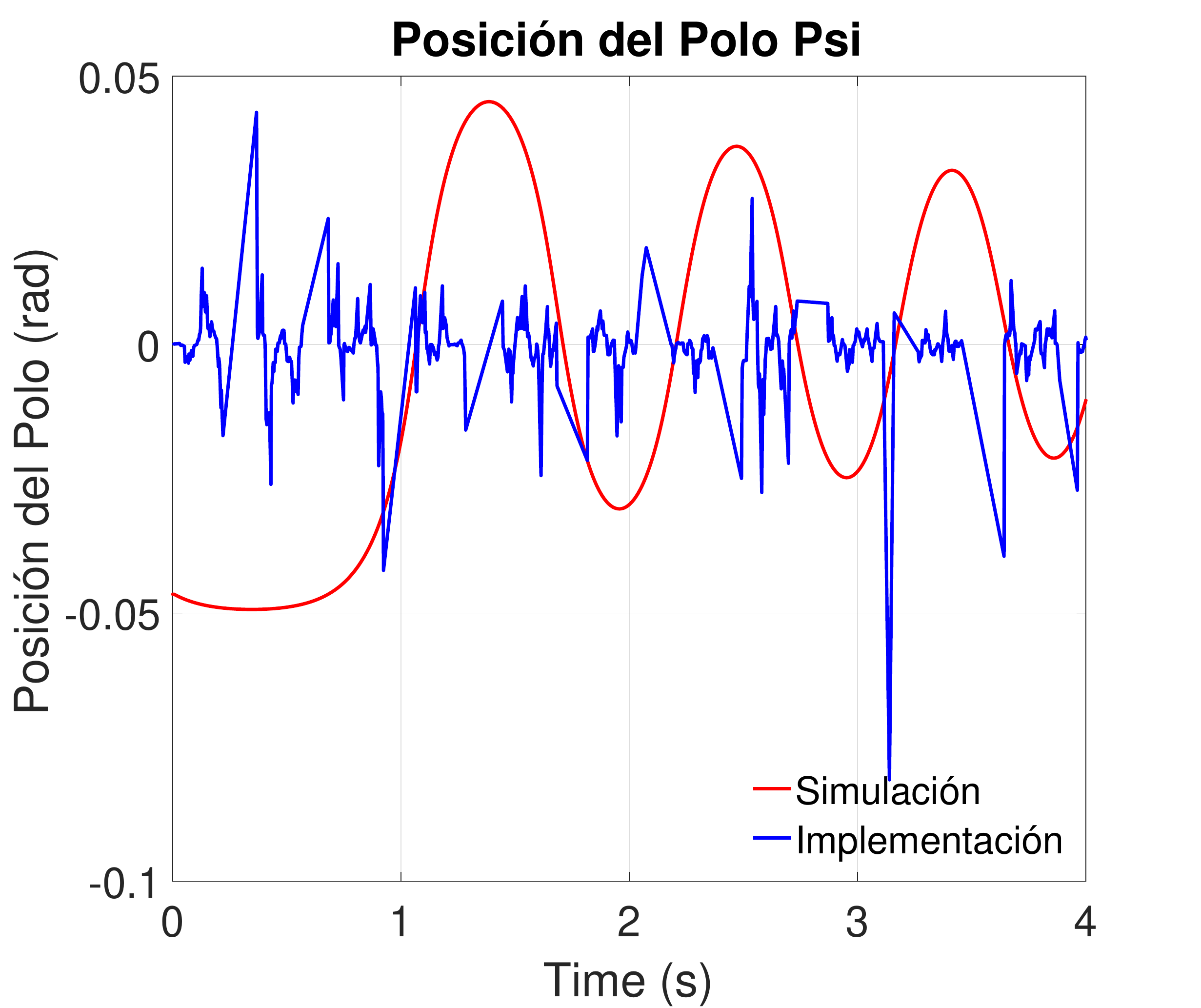}}
	\subfigure[NxtWayPsiVel SMC]{
		\label{f:nxtwayapsmc}
		\includegraphics[width=0.23\textwidth]{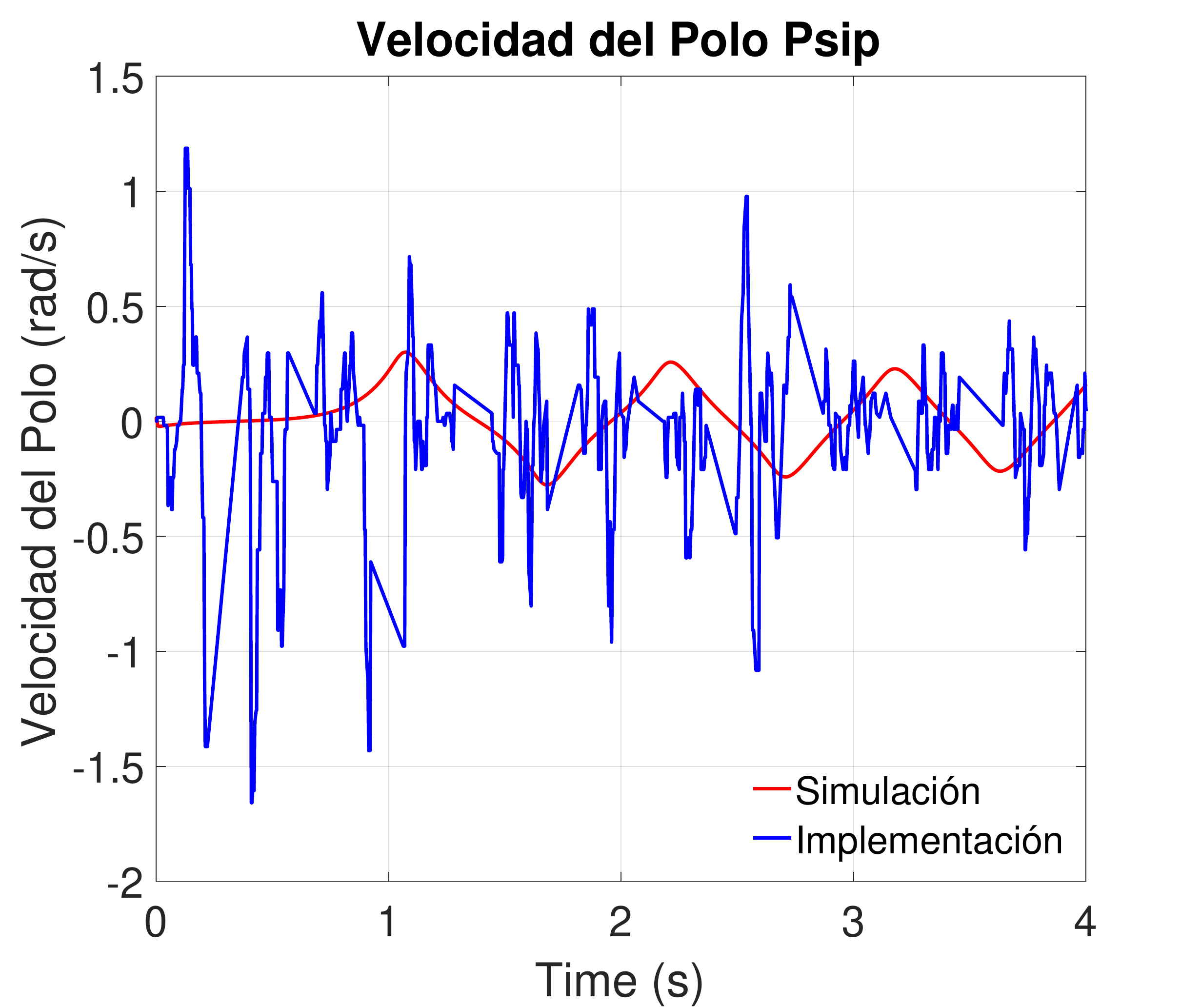}}
	\subfigure[NxtWaySalida SMC]{
		\label{f:NxtwayUsmc}
		\includegraphics[width=0.23\textwidth]{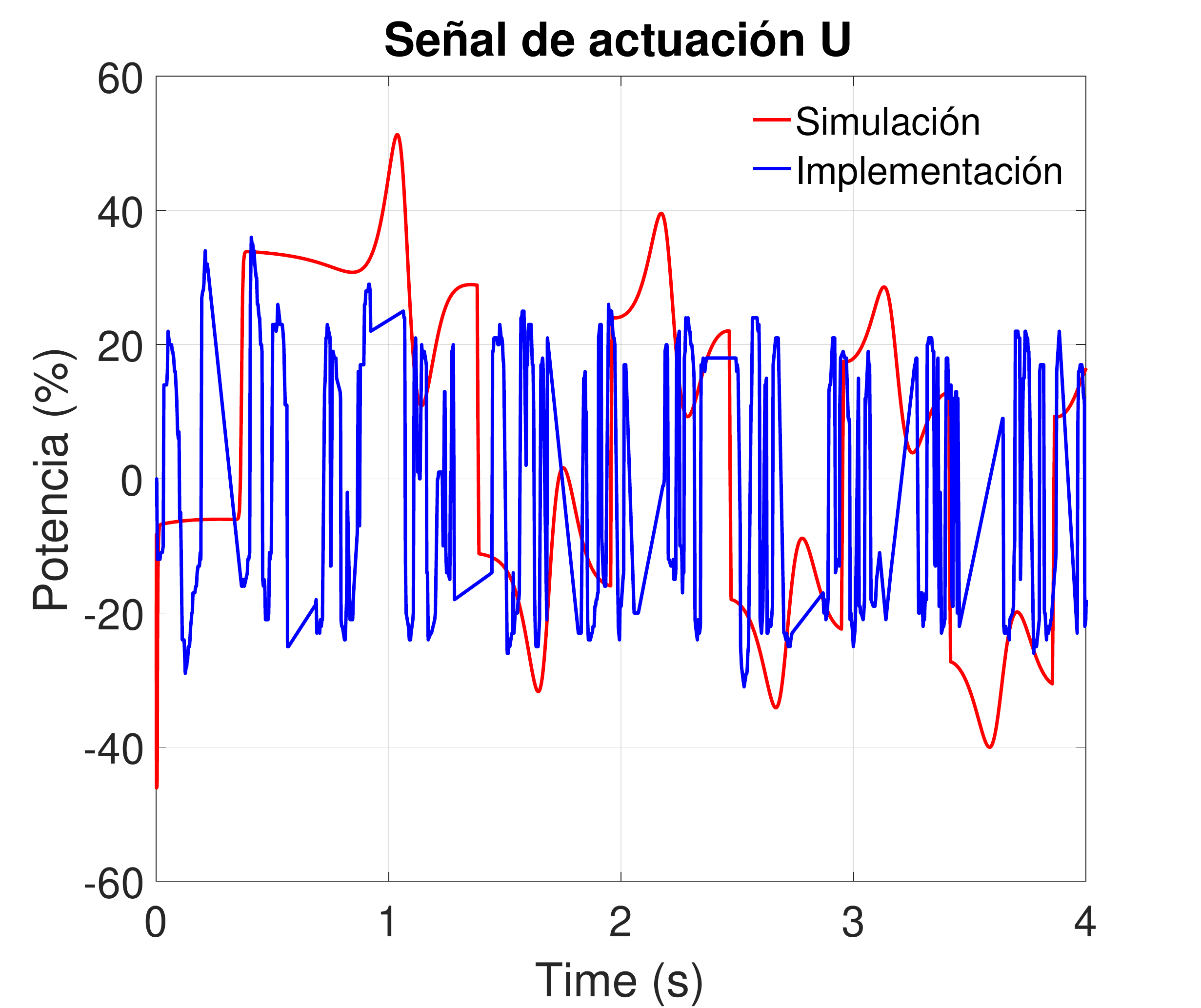}}
	\caption{Respuesta control SMC Péndulo NxtWay}
	\label{fig:smcn}
\end{figure}

\begin{figure}[htbp]
	\centering
	\subfigure[RotPenThetaPos SMC]{
		\label{f:RotPentsmc}
		\includegraphics[width=0.23\textwidth]{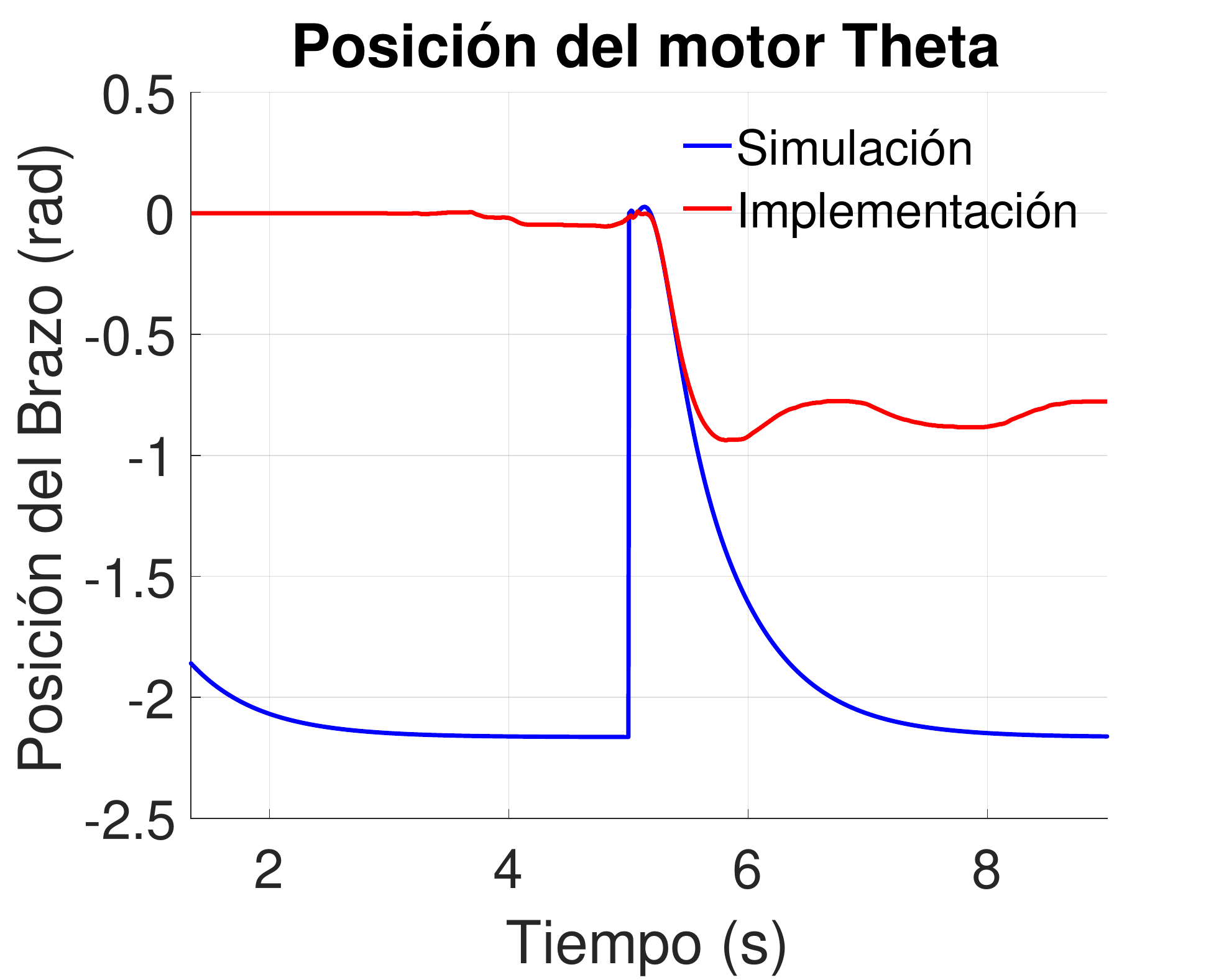}}
	\subfigure[RotPenThetaVel SMC]{
		\label{f:RotPentpsmc}
		\includegraphics[width=0.23\textwidth]{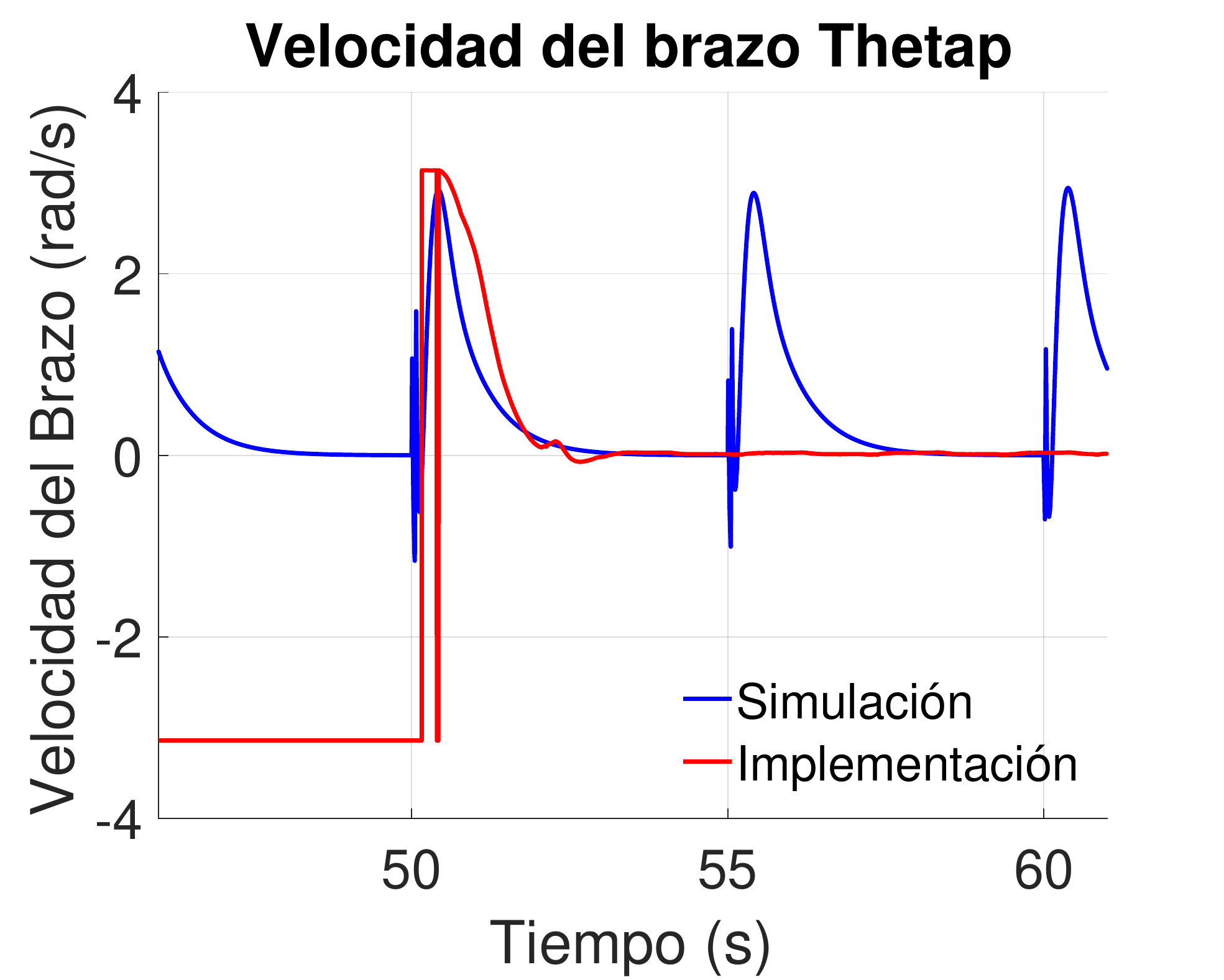}}
	\subfigure[RotPenPsi SMC]{
		\label{f:RotPenasmc}
		\includegraphics[width=0.23\textwidth]{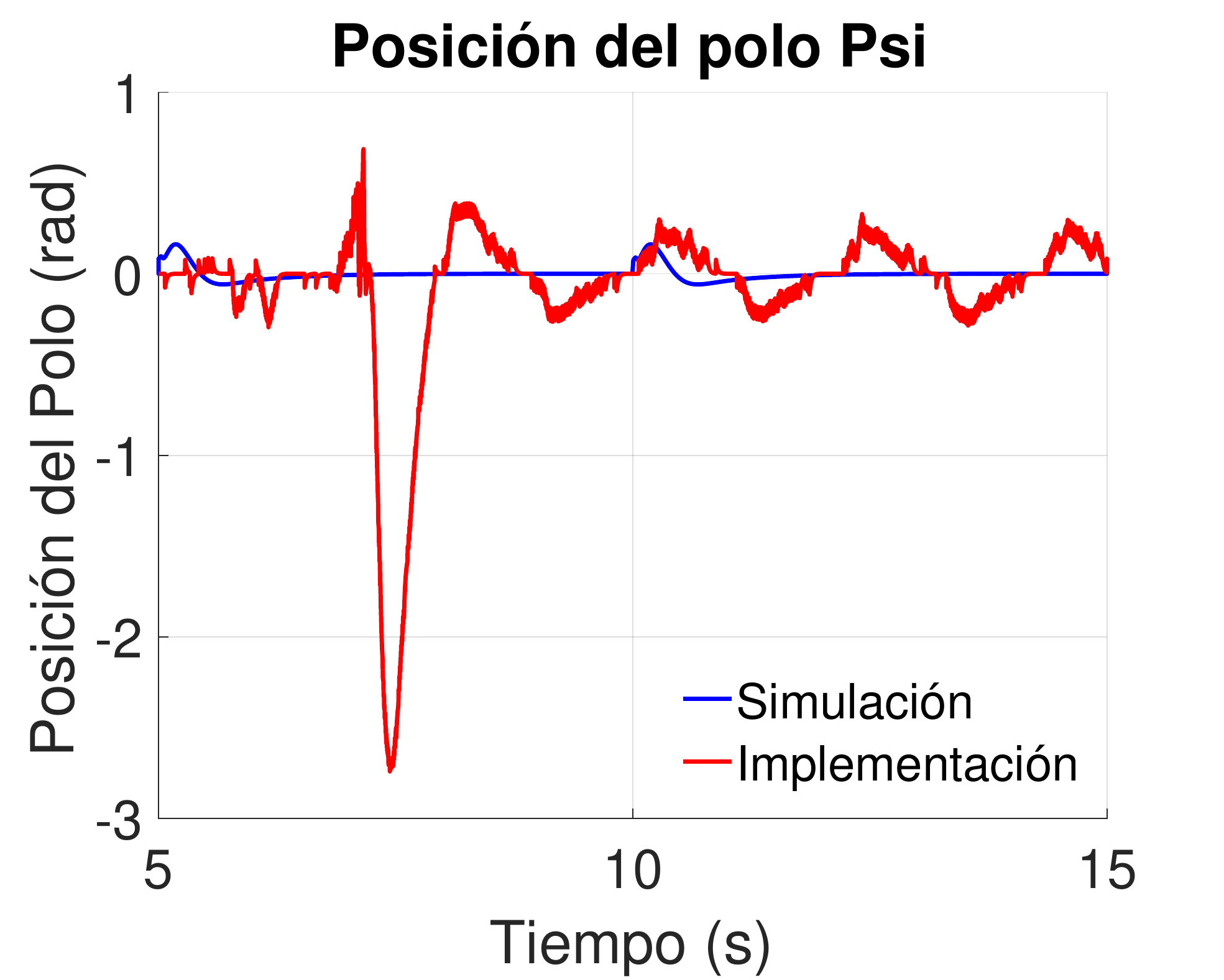}}
	\subfigure[RotPenPsiVel SMC]{
		\label{f:RotPenapsmc}
		\includegraphics[width=0.23\textwidth]{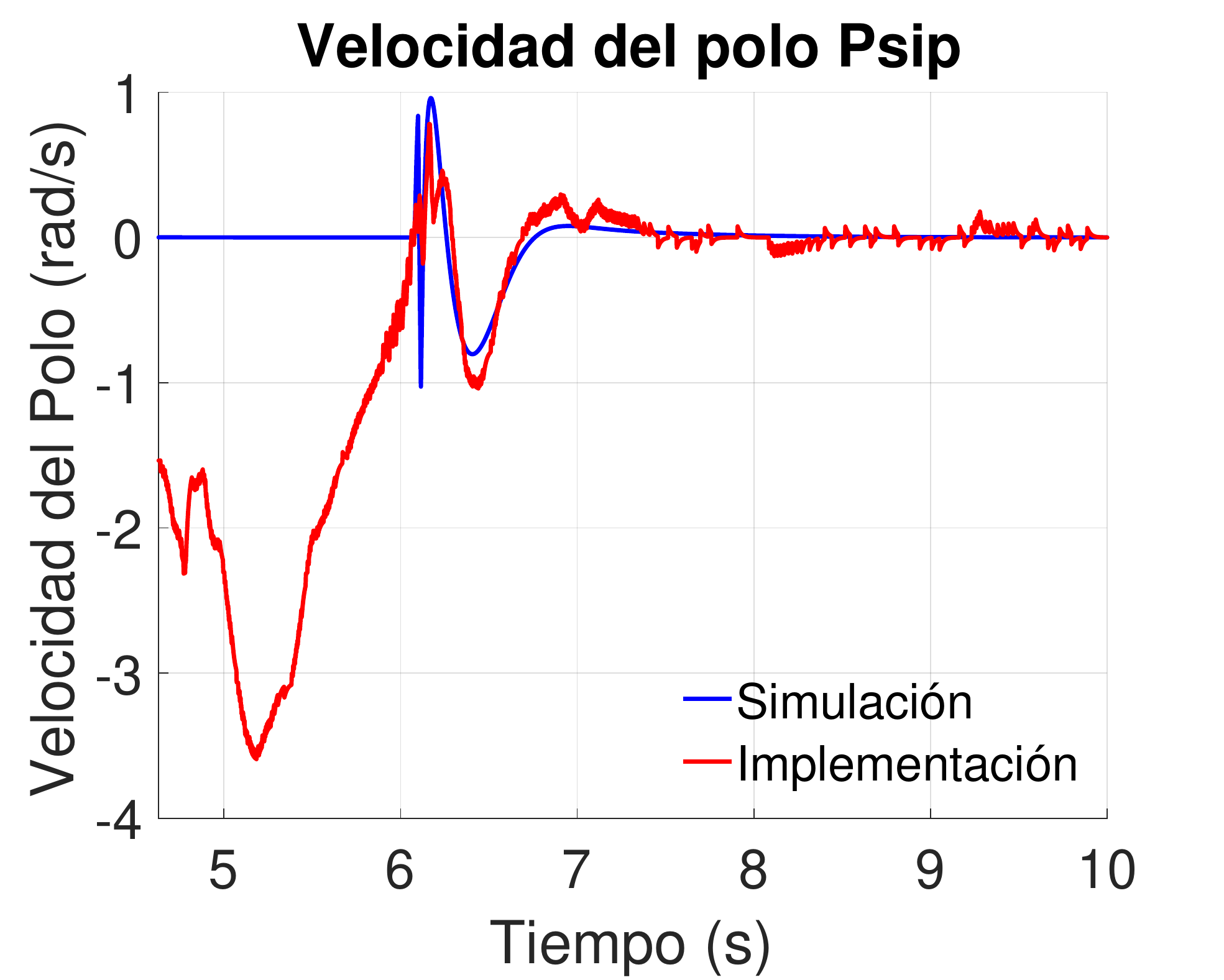}}
	\subfigure[RotPenSalida SMC]{
		\label{f:RotPenUsmc}
		\includegraphics[width=0.23\textwidth]{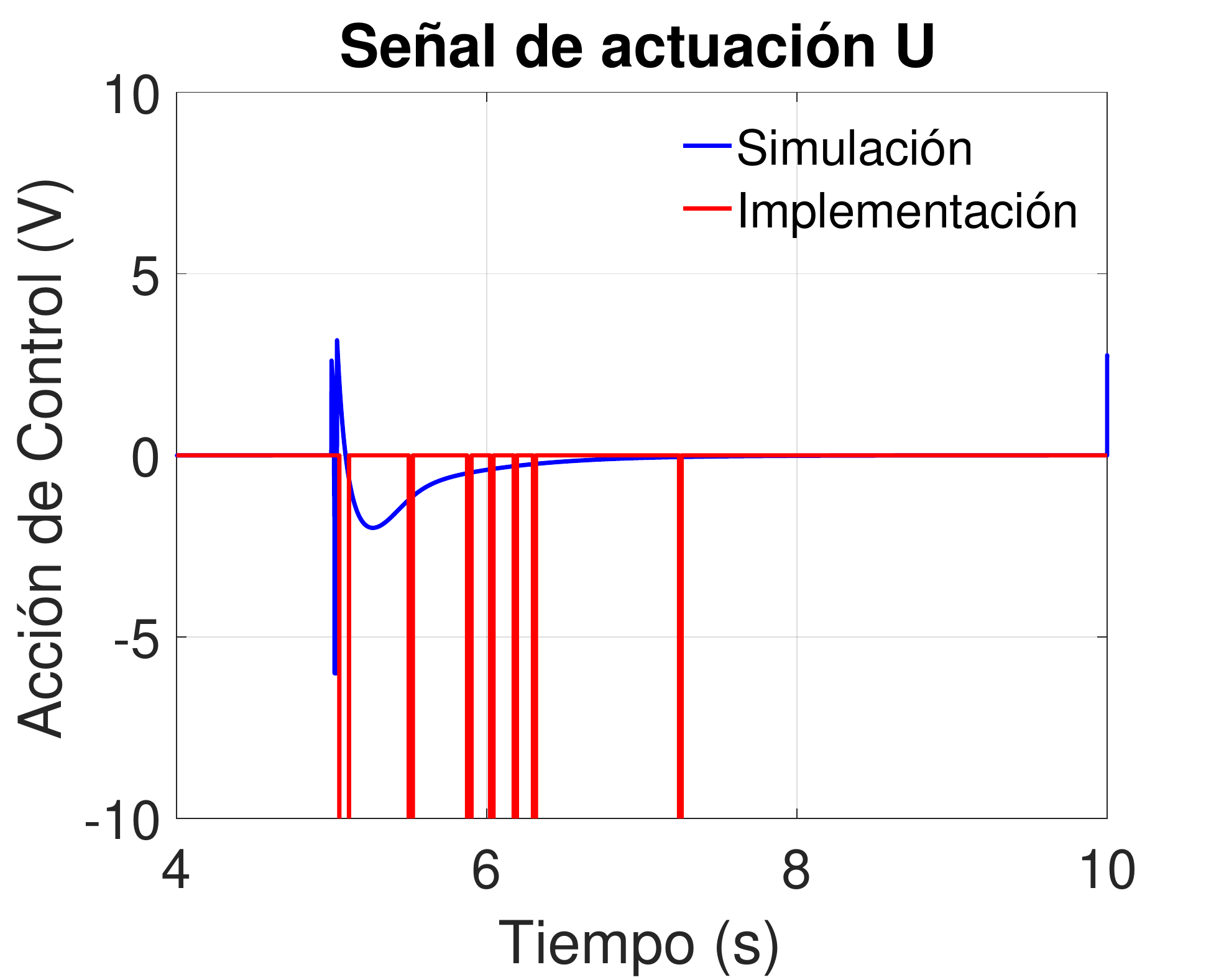}}
	\caption{Respuesta control SMC Péndulo Quanser}
	\label{fig:smcq}
\end{figure}

A partir de la respuesta en lazo cerrado de cada uno de los sistemas controlados, se lleva a cabo una comparación de rendimiento bajo índices de calidad; Los índices de validación son: nivel de estabilización de la variable de estado asociada al polo, energía de la señal de control (Norma infinito de la señal), máxima velocidad de la variable asociada al polo, y consideraciones de síntesis de controladores cuando se procede al diseño. Evidenciando estos valores en las figuras \ref{fig:fig5}, \ref{fig:fig6}, \ref{fig:smcn} y \ref{fig:smcq}; se observa que ambas plataformas logran la estabilización de sus estructuras ante una perturbación. Para resumir algunos resultados de importancia para la aplicación, se proponen las siguientes tablas resumen. 

\begin{table}[htbp]
	\centering
	\caption{Desempeño NXT}
	\label{Tabla_comp1}
	\resizebox{.5\textwidth}{!}{
		\begin{tabular}{llllcl}
			\cline{2-6}
			\multicolumn{1}{l|}{}&\multicolumn{1}{c|}{\begin{tabular}[c]{@{}c@{}}Estado $q_2$\\ 
					Nivel\\
					Estabilización\end{tabular}}&\multicolumn{1}{c|}{\begin{tabular}[c]{@{}c@{}}Potencia\%\\
					Potencia\\
					Máxima\end{tabular}}&\multicolumn{1}{c|}{\begin{tabular}[c]{@{}c@{}}$\dot{q_2}$\\
					Velocidad\\
					M\'axima\end{tabular}}&\multicolumn{1}{c|}{\begin{tabular}[c]{@{}c@{}}Criterio\\
					Energía\\
					Mínima\end{tabular}}&\multicolumn{1}{l|}{Robustez}\\
			\hline
			\multicolumn{1}{|l|}{LQR}&\multicolumn{1}{l|}{Acotado(Suave)}& \multicolumn{1}{l|}{5\%}&\multicolumn{1}{l|}{1.6}&\multicolumn{1}{c|}{Si}&\multicolumn{1}{c|}{No}\\
			\hline
			\multicolumn{1}{|l|}{SMC}&\multicolumn{1}{l|}{Acotado (Scattering)}& \multicolumn{1}{l|}{27\%}&\multicolumn{1}{l|}{1.2}& \multicolumn{1}{c|}{No}&\multicolumn{1}{c|}{Si}\\
			\hline& & & &\multicolumn{1}{l}{} &
		\end{tabular}
	}
\end{table}

En la tabla \ref{Tabla_comp1} se observan métricas en las variables de relevancia del sistema para medir desempeño de los dos controladores en el módulo NXT. Esta información puede ser usada para determinar un criterio de selección en aplicaciones futuras, por ejemplo, en cuanto al uso energético es mejor usar un controlador LQR que uno basado en SMC, esto se puede inferir a partir de la consideración de la máxima potencia aplicada en el servomotor y el criterio de energía usado en la sintetización del controlador.

En la tabla \ref{Tabla_comp2}, a diferencia del caso anterior, se tiene la comparación de técnicas en el sistema Quanser, mostrando un comportamiento mejorado por la suavidad de sus señales. Si se considera una selección basada en la robustez de la técnica, es mejor seleccionar SMC, debido a que en la sintetización es posible incluir perturbaciones acotadas, lo cual reformula la desigualdad de Lyapunov, permitiendo recalcular las constantes de realimentación y lograr estabilidad en tiempo finito.

\begin{table}[htbp]
	\centering
	\caption{Desempeño Quanser}
	\label{Tabla_comp2}
	\resizebox{.5\textwidth}{!}{
		\begin{tabular}{llllcl}
			\cline{2-6}
			\multicolumn{1}{l|}{}&\multicolumn{1}{c|}{\begin{tabular}[c]{@{}c@{}}Estado $q_2$\\
					Nivel\\
					Estabilización \end{tabular}} &\multicolumn{1}{c|}{\begin{tabular}[c]{@{}c@{}}Potencia (V)\\ Potencia\\Máxima\end{tabular}}&\multicolumn{1}{c|}{\begin{tabular}[c]{@{}c@{}}$\dot{q_2}$\\
					Velocidad\\ Maxima\end{tabular}}&\multicolumn{1}{c|}{\begin{tabular}[c]{@{}c@{}}Criterio\\
					Energía\\
					Mínima\end{tabular}} & \multicolumn{1}{l|}{Robustez}\\
			\hline
			\multicolumn{1}{|l|}{LQR}&\multicolumn{1}{l|}{Acotado (Suave)}& \multicolumn{1}{l|}{3.6}&\multicolumn{1}{l|}{1.5}& \multicolumn{1}{c|}{Si}&\multicolumn{1}{c|}{No}\\
			\hline
			\multicolumn{1}{|l|}{SMC}&\multicolumn{1}{l|}{Acotado (Bajo Scattering)}&\multicolumn{1}{l|}{10}&\multicolumn{1}{l|}{3.6}&\multicolumn{1}{c|}{No}&\multicolumn{1}{c|}{Si}\\
			\hline
			& & & & \multicolumn{1}{l}{} &                              
		\end{tabular}
	}
\end{table}

Es importante destacar que las gráficas visualizadas no muestran una relación directa entre implementación y simulación, debido a que en los cálculos no se maneja la totalidad de las variables físicas involucradas en el modelamiento de las plataformas; y así mismo, las variables involucradas, poseen un cierto grado de incertidumbre asociado. El uso de un controlador óptimo, permite mediante criterios de optimización, llevar los polos del sistema a una región estable, sin llegar a seleccionarlos como ocurre con el control en modo deslizante, lo que permite tener un mejor proceso de diseño en el cálculo de sus constantes; por otro lado, el control en modo deslizante permite incluir en sus dinámicas y metodologías de diseño, perturbaciones externas al sistema, sin afectar las constantes a obtener, a diferencia del control LQR, en donde el diseño se realiza unicamente sobre el modelo de estados de la planta, e incluir estas dinámicas, harían necesario reiniciar el proceso nuevamente, perdiendo robustez en su diseño.

\section{Conclusiones}
En este trabajo, fueron seleccionados dos estructuras pendulares disponibles para experimentación. Inicialmente, se realizó el modelo dinámico de las plataformas usando las ecuaciones de Euler-Lagrange, cada modelo fue validado mediante el uso de herramientas computacionales; ambos sistemas tuvieron una respuesta similar en cuanto a la posición del polo, la posición del brazo por estar restringida físicamente en la plataforma RotPen, describe una dinámica y respuesta diferente.

A partir de los modelos y de su linealización, se desarrollaron controladores LQR y SMC, cada uno de estos fue probado en Simulink. Además, se implementaron los controladores en los entornos de cada fabricante, el péndulo rotatorio RotPen se implementó en la herramienta de Simulink de Matlab, mientras que el péndulo móvil NxtWay, mediante texto estructurado, se realizó en el software RobotC. En cada caso se añadió una perturbación para obtener los resultados donde se midió la norma infinito de la señal de control, la calidad de la estabilización, y la norma infinito de la velocidad del polo. La perturbación consistió en una señal sinusoidal de pulsos, con un voltaje pico equivalente a la mitad de la potencia de cada actuador a una frecuencia de 0.0167Hz. Esta estrategia permitió validar la calidad de funcionamiento del control; tomando en cuenta las restricciones de hardware y software que cada dispositivo poseía; no obstante, se observa que sin importar estas restricciones, el diseño de un controlador óptimo y uno basado en modos deslizantes, son estrategias adecuadas para la estabilización de estructuras pendulares, evidencia que se puede observar en las gráficas de funcionamiento de cada dispositivo con cada uno de los controladores empleados.

A modo de comparación, se observó un mejor rendimiento en el péndulo móvil NxtWay, por la simplicidad de su entorno de desarrollo para embeber los algoritmos de control, esto permitió una mayor velocidad en la ejecución de una inferencia de control, lo que llevó a tener una respuesta más rápida ante perturbaciones, cuantificado en un segundo aproximadamente para la plataforma Rotpen, y en 1.5 segundos para la plataforma NxtWay; tomando en cuenta que la plataforma RotPen empleaba una frecuencia de procesamiento de 500 inferencias por segundo, mientras que la plataforma NxtWay presentaba una frecuencia de 250 inferencias por segundo.

Un aspecto a tener en cuenta para la implementación de controladores en esta clase de dispositivos, es la respuesta en alta frecuencia de los sensores, la cual afecta el cálculo de las derivadas e integrales de las señales involucradas, al adicionar componentes de ruido que enmascaran la correcta lectura de las variables de estado del sistema; por esta razón se hace necesario el diseño de filtros pasa-bajos que eliminen estas componentes y permitan una lectura adecuada de las variables.

\section*{Trabajo futuro}

Como trabajo futuro, se plantea la identificación de los sistemas controlados, con el fin de implementar una técnica de control no lineal, que aumente la región de operación de los dispositivos, sin disminuir su rendimiento ni incrementar su gasto computacional; la técnica de identificación seleccionada es la técnica difusa Takagi-Sugeno, la cual se ajusta a una propuesta de controlador no lineal que actúa como múltiples controladores en paralelo ponderados no linealmente \cite{takagi1985fuzzy}. Esta identificación se comparará con la versión no lineal del controlador SMC.


\section*{Agradecimientos}

Los autores agradecen al laboratorio de la facultad de Ingeniería en Automatización de La universidad de La Salle por apoyo técnico y logístico, a la Vicerrectoría de Investigación y Transferencia Salle, y a Colciencias por financiar este proyecto a través de las convocatorias VRIT 2432 y Jovénes Investigadores e Innovadores Colciencias 761 de 2016. Se hace un reconocimiento adicional nuevamente a Colciencias, por la beca doctoral Francisco José de Caldas Generación Bicentenario de 2009 otorgada a C. H. Rodriguez-Garavito, de donde se derivó este proyecto de investigación.

\bibliographystyle{IEEEtran}
\bibliography{main.bib}

\end{document}